\documentclass[twocolumn]{aa} 

\usepackage{graphicx}
\usepackage{txfonts}

\begin{document}

\def\deg{$^{\rm o}$}

\title{Dying Radio Galaxies in Clusters}

\author{M. Murgia \inst{1} \and P. Parma\inst{2} \and K.-H. Mack\inst{2} 
\and H.R. de Ruiter\inst{2} \and R. Fanti\inst{2} \and F. Govoni\inst{1} \and A. Tarchi\inst{1} \and S. Giacintucci\inst{3} \and M. Markevitch\inst{3} }

\institute{
INAF\,-\,Osservatorio Astronomico di Cagliari, Loc. Poggio dei Pini, Strada 54,
I-09012 Capoterra (CA), Italy
\and
INAF\,-\,Istituto di Radioastronomia, Via Gobetti 101, I-40129 Bologna, Italy
\and
Harvard-Smithsonian Center for Astrophysics, 60 Garden St., Cambridge, MA 02138, USA 
}

\date{Received; Accepted}

\abstract
{}
{In this work we present the study of five `dying' nearby ($z\le0.2$) radio galaxies belonging to the WENSS minisurvey and to the B2 bright 
catalogs: WNB1734+6407, WNB1829+6911, WNB1851+5707, B2\,0120+33, and B2\,1610+29.}
{These sources have been selected on the basis of their extremely steep broad-band radio spectra, which
is a strong indication that these objects either belong to the rare class of dying radio galaxies or 
that we are  observing `fossil' radio plasma remaining from a previous nuclear activity. We derive the relative duration of 
the dying phase from the fit of a synchrotron radiative model to the radio spectra of the sources.}
{The modeling of the integrated spectra and the deep spectral index images obtained with the VLA 
 confirmed that in these sources the central engine has ceased to be active for a significant fraction of their lifetime although their extended lobes have
not yet completely faded away. We found that WNB1851+5707 is in reality composed by two distinct dying galaxies, which appear blend together as a single source
in the WENSS. In the cases of  WNB1829+6911 and  B2 0120+33, the fossil radio lobes are seen in conjunction
with a currently active core. A very faint core is detected also in a MERLIN image of WNB1851+5707a,
 one of the two dying sources composing WNB1851+5707. We found that all sources of our sample are located (at least in projection) at the center of an X-ray emitting cluster.}
{Our results suggest that the duration of the dying phase for a radio source in
cluster can be significantly higher with respect to that of a radio galaxy in the field, although no firm conclusions can be drawn due to the small number
 statistics involved. The simplest interpretation for the tendency of dying galaxies to be found in cluster is that the low-frequency 
radio emission from the fading radio lobes last longer if their expansion is somewhat reduced or even stopped. 
Another possibility is that the occurrence of dying sources is higher in galaxy clusters. We argue that radio sources in 
dense environment, like e.g. the center of cooling core clusters, may
have a peculiar accretion mode which results in a bursting duty cycle sequence of active and quiescent periods. 
This result could have important implications for theories of the life cycles of radio sources and AGN feedback in clusters of galaxies but awaits
confirmation from future observations of larger, statistically significant, samples of objects.}


\offprints{M. Murgia, matteo@oa-cagliari.inaf.it}

\maketitle

\section{Introduction}

Dying radio galaxies represent an interesting, but still largely
unexplored, stage of the active galactic nuclei evolution.
During their active stage, which may last several $10^{7}$ years, the strong 
radio sources associated with elliptical galaxies are supplied with
energy from active galactic nuclei via plasma beams or jets. Due to the continuous
 accumulation of new particles, the total spectra of the active radio sources are usually well 
approximated by a power law over a wide range of frequencies.
The injection of energy also sustains the growth of these radio sources which 
is governed by the balance between the internal pressure in the radio
lobes and the pressure in the hot X-ray emitting external medium in which they must expand (Scheuer 1974).

At some point, however, the activity in the nuclei stops or falls 
to such a low level that the plasma outflow can no longer be sustained and the radio source is 
expected to undergo a period of fading (dying phase) before it disappears completely.
In the dying phase, radio core, well-defined jets and compact hot-spots will disappear because 
they are the structures produced by continuing activity. On the other hand, the radio lobes 
may still remain detectable for a long time if they are subject only to radiative losses
of the relativistic electrons. The first example of such sources is B2 0924+30 and was illustrated by Cordey 
(1987). It is also possible that radio galaxies
may be active intermittently or even that jets flicker before eventually going off completely. 
In this scenario, one expects to observe
fossil radio plasma remaining from an earlier active epoch, along with newly
restarting jets. The best case for fossil radio lobes seen with a currently active galaxy
is 3C 338 (see Gentile et al. 2007 for a recent work). The very steep spectrum lobes of this source
 are clearly disconnected from the currently active jets.

\begin{table*}[t]
  \begin{center}
  \caption{Basic properties for the dying radio galaxies candidates. Spectral indexes 
and flux densities are from the WENSS and NVSS. The spectral curvatures are calculated by
measuring $\alpha_{\rm low}$ between 151 and 325 MHz, for the WENSS minisurvey sources, and between 
151 and 408 MHz, for the B2 sources. The high-frequency spectral index  $\alpha_{\rm high}$ is taken between 1400 and 4850 MHz
for all sources.}

\medskip
  \label{tab:pages}
  \begin{tabular}{ccccccccc}
\hline
\noalign{\smallskip}
    Source & R.A.\,(J2000) & DEC.\,(J2000) & redshift & scale & $S_{325 \rm MHz}$  & $\alpha_{1.4 \rm GHz}^{325\rm MHz}$ & $SPC$\\
           & (h m s)       & (d m s)       &          & (kpc/\arcsec)&   (mJy)             &        & \\ 
\hline
\noalign{\smallskip}
 WNB1734+6407   & 17 35 04.6 & +64 06 07.7 & 0.1406  &  2.4    & 120  &  2.0 & 1.1 \\
 WNB1829+6911   & 18 29 05.6 & +69 14 06.0 & 0.204   &  3.3    & 209  &  1.9 & -0.4\\
 WNB1851+5707a  & 18 52 08.5 & +57 11 42.3 & 0.1068  &  1.9    & 368  &  1.7 & 1.1\\
 WNB1851+5707b  & 18 52 09.7 & +57 11 56.6 & 0.1066  &         &      &      &   \\
 B2 0120+33     & 01 23 39.9 & +33 15 22.1 & 0.0164  & 0.33    & 727  &  1.3 &1.5\\
 B2 1610+29     & 16 12 35.5 & +29 29 05.3 & 0.0318  & 0.63    & 412  &  0.9 &0.8\\
\hline
\end{tabular}
\end{center}
\end{table*}

In this work we classify a radio
source as dying not only if the fossil lobes are detached from the AGN and there is no evidence for nuclear activity (trivial case), 
but also if some kind of nuclear activity is present but the fossil lobes still dominate the total source's radio luminosity\footnote{
This choice is motivated by the following two considerations. First, even if the nuclear component we are observing is really
produced by new couple of restarting radio jets, we would never know a priori if this new activity would be able 
to last enough to regenerate a radio galaxy similar to previous event of radio activity whose
the fossil lobes represent the remains. Second, it could not be obvious to recognize if a nuclear component can be considered as the
begin of a new phase of AGN activity or rather a fading radio core which is going-off definitely.}. Hence, in what follows we will 
refer indifferently to dying and re-starting sources if the fossil lobes are the dominant component while we shall eventually
reserve the term \emph{renewed source} to indicate the case in which the new and the fossil components are roughly comparable in
luminosity, like e.g. in the case of the double-double radio sources (Schoenmakers et al. 2000; Saripalli et al. 2002, 2003).

Indeed, given the comparatively short duration of the radio galaxy phenomenon, we could 
expect a large number of dying radio sources.
However, only a handful of dying radio galaxies in
this evolutionary stage are known. According to Giovannini et al. (1988) only few percent 
of the radio sources in the B2 and 3C samples have the characteristics of a dying radio galaxy. It 
is important to note that these sources
represent the last phase in the `life' of a radio galaxy. Therefore they must be well distinguished from
the cluster radio halo/relic phenomenon, that is usually not associated with an
 individual galaxy (see Ferrari et al. 2008 and references therein).

A possible explanation for the rarity of the dying radio galaxies may be the
relatively fast spectral evolution they undergo during the fading phase. Synchrotron losses and
the inverse Compton scattering of the Cosmic Microwave Background photons
preferentially deplete the high-energy electrons.  The fading lobes are expected to have
very steep ($\alpha > 1.3$; $S_{\rm \nu} \propto\nu^{-\alpha}$) and convex
radio spectra characteristic of a population of electrons which have
radiated away much of their original energy (Komissarov \& Gubanov 1994).
In fact, in the absence of fresh particle injection, the high-frequency radio spectrum develops an
exponential cutoff. At this point, the adiabatic expansion of the
 radio lobes will concur to shift this spectral break to lower frequencies
 and the source will disappear quickly. On the other hand, if the source expansion
is somehow reduced, or even stopped, there is still the chance to detect the
fossil radio lobe, at least at low frequency.

For the reasons mentioned above, low-frequency selected samples such us the Westerbork 
Northern Sky Survey (WENSS; Rengelink et al. 1997) at 325 MHz and the B2 survey at 408\,MHz (Colla et al. 1970, 1972, 1973) are particularly
well-suited to search for these elusive fossil radio sources. Parma et al. (2007),
by cross-correlating the WENSS with the NVSS (Condon et al. 1998), discovered six new dying sources
and three new restarted sources. Other two dying galaxies (the central radio source in Abell 2622 and MKW03s) and 
 one possibly restarting source (MKW07) have been recently found by Giacintucci et al. (2007) in a 
low-frequency survey of nearby cluster of galaxies performed with Giant Metrewave Radio Telescope. 

In this work we report on the result of a search for dying sources in
a WENSS sub-sample (the minisurvey; de Ruiter et al. 1998) and in the B2 bright sample described
 by Colla et al. (1975). Both samples have been constructed from the parent surveys by
 identifying the radio sources with nearby bright galaxies.
From these samples we have selected five dying radio galaxies candidates whose extremely steep spectra are characterized by
a quasi-exponential drop above a frequency of about 1 GHz. These are WNB1734+6407, WNB1829+6911, WNB1851+5707, B2 0120+33, 
and B2 1610+29.

In order to determine whether these sources were really dying objects or
fossil lobes associated with restarting radio galaxies,
we studied their radio continuum emission with the Very Large Array (VLA) at
various frequencies. The new observations and the analysis of the data already available in literature 
permitted us to derive the detailed integrated radio spectra and spatially resolved
 spectral index images for all the five candidates. We also had the 
chance to observe at sub-kpc resolution the sources WNB 1851+5707 with the Multi-Element Radio
 Linked Interferometer Network (MERLIN) interferometer.

Furthermore, we found that all the five sources are unambiguously associated 
to a diffuse X-ray emission in the  Rosat All Sky Survey (RASS). 
In all cases the association is with a known cluster of galaxies.

In this work we report the results of this extensive campaign of radio observations.
In Sect.\,2 we described the criteria of the source selection. The descriptions of the VLA and MERLIN 
observations are reported in Sect.\,3 and 4, respectively. In Sect.\,5 we present the analysis of the
 radio spectra. The X-ray environment is presented in Sect.\,6. Finally, a summary of the work is given in Sect.\,7.

Throughout this paper we assume a $\Lambda$CDM cosmology with
$H_0$ = 71 km s$^{-1}$Mpc$^{-1}$, $\Omega_m$ = 0.3, and $\Omega_{\Lambda}$ = 0.7.

\section{Source selection}
\subsection{The spectral curvature of dying and restarting sources}
In the active stage the total spectra of the radio galaxies are usually
well approximated by a power law over a wide range of frequencies. Spectral
breaks at high frequencies are also often observed. The break frequency is due to the
radiative losses of the synchrotron electrons and it can be related
to the magnetic field and age of the source. If the integrated spectrum of the radio
source is dominated by the emission of the radio lobes, which accumulated the electrons injected by the jets, 
it can be modeled by a continuous injection model (CI) in which the low-frequency spectral
 index $\alpha_{\rm low}$ represents the injection spectral index of the youngest electron populations, 
 $\alpha_{\rm inj}$, while the high-frequency spectral index is limited to 
$\alpha_{\rm high}\le\alpha_{\rm inj}+0.5$ at frequencies greater than the break frequency (see e.g. Murgia et al. 1999).
As the source ages, the break moves to low frequency and the radio spectrum eventually becomes a steep power law in which 
$\alpha_{\rm low}$ approaches $\alpha_{\rm high}$.
For dying sources however we expect an exponential cut-off of the integrated spectrum above the break frequency (e.g. Parma et 
 al. 2007). In this case the high-frequency spectral index, $\alpha_{\rm high}$, may became much steeper than the low-frequency one, $\alpha_{\rm low}$. On 
the other hand, if the activity in the galaxy nucleus restarts $\alpha_{\rm high}$ can be flatter that $\alpha_{\rm low}$. This suggests a
 possible selection criterion based on the radio spectrum curvature to separate active, dying and restarting sources.
Following the idea of Sohn et al. (2003) we define the spectral curvature\footnote{Please note that Sohn et al. 2003 adopted a
 slightly different definition for $SPC$ with respect to the one proposed here.}, $SPC$, as
\begin{equation}
SPC=\alpha_{\rm high}-\alpha_{\rm low}
\end{equation} 
As far as the optically thin portion of the radio spectrum is considered, in the framework of the CI model the spectral curvature will be very close to $SPC\approx 0$. In fact, the maximum curvature we could expect in the active phase is limited to $\alpha_{\rm high}-\alpha_{\rm low}\le 0.5$.

On other hand, the integrated spectra of dying sources are characterized by an exponential cut-off at high frequencies and thus for them extreme spectral curvatures ($SPC\gg 0.5$) are possible. Finally, for restarting sources the injection of new electrons will flatten the high frequency spectrum resulting in $\alpha_{\rm high}<\alpha_{\rm low}$, and thus $SPC<0$ (Murgia, in preparation). 

Obviously, the proposed spectral curvature classification must be always considered to have a statistical meaning and used by taking into account
 the following caveats and warnings. 
The $SPC$ to be meaningful requires to be calculated over a broad frequency window. Because of the nature of the synchrotron radiation, 
the spectral curvature of integrated spectra generally occurs over a frequency range of at least two order of magnitudes. 
Hence, if the observed spectral window is too small, there is always the possibility to miss-classify as active a radio galaxy which is indeed
 dying. Re-starting sources are particularly problematic from this point of view since, as the injection of new radio plasma proceeds, the spectral
curvature of the source eventually decreases from $SPC>0.5$ to $SPC<0$, so all values of $SPC$ are in principle possible for them (see Sect. 2.3 for the case of 3C 338).
Moreover, individual radio sources could not fit in the above picture if their 
integrated spectrum is dominated by  Doppler boosted components, like jets, flat spectrum cores, etc. Other common problems in radio surveys, like e.g. 
the blending of unrelated sources or the difficulties in imaging extended sources, could lead to spurious $SPC$.
For all these reasons, the final confirmation of a dying source necessarily requires the follow-up with deep, spatially resolved, spectral index radio imaging of the dying candidate. 

With this strategy in mind ($SPC$ selection and deep spectral index imaging) we searched for dying and restarting sources candidates in two
 samples extracted from the WENSS and the B2 catalogs.

\subsection{The WENSS minisurvey sample}
Dying radio galaxies are more easily detected at low frequency, therefore
the WENSS at 325 MHz is particularly well-suited to search for these elusive objects.
We are interested in steep spectrum radio sources identified with bright galaxies as these 
should be prime candidates for dying sources (Parma et al. 2007). 
The WENSS minisurvey contains 402 radio sources identified with either elliptical or spiral galaxies 
 with red magnitude brighter than roughly $m_{\rm r}= 17 - 18$.
Using cross-correlation with the existing catalogs and
new observations with the Effelsberg 100-m telescope at 4.8 and 10.4\,GHz (Mack et al. in preparation), 
we have obtained spectral information in the frequency
range from 38 MHz to 10 GHz for about 200 sources brighter than $S_{325}>30$.  
The spectral curvature can be used to select dying radio galaxies {\it candidates} in large samples, like the minisurvey, where detailed high-resolution images 
are not available but there is a good spectral coverage.

\begin{figure*}
\begin{center}

\includegraphics[width=18cm]{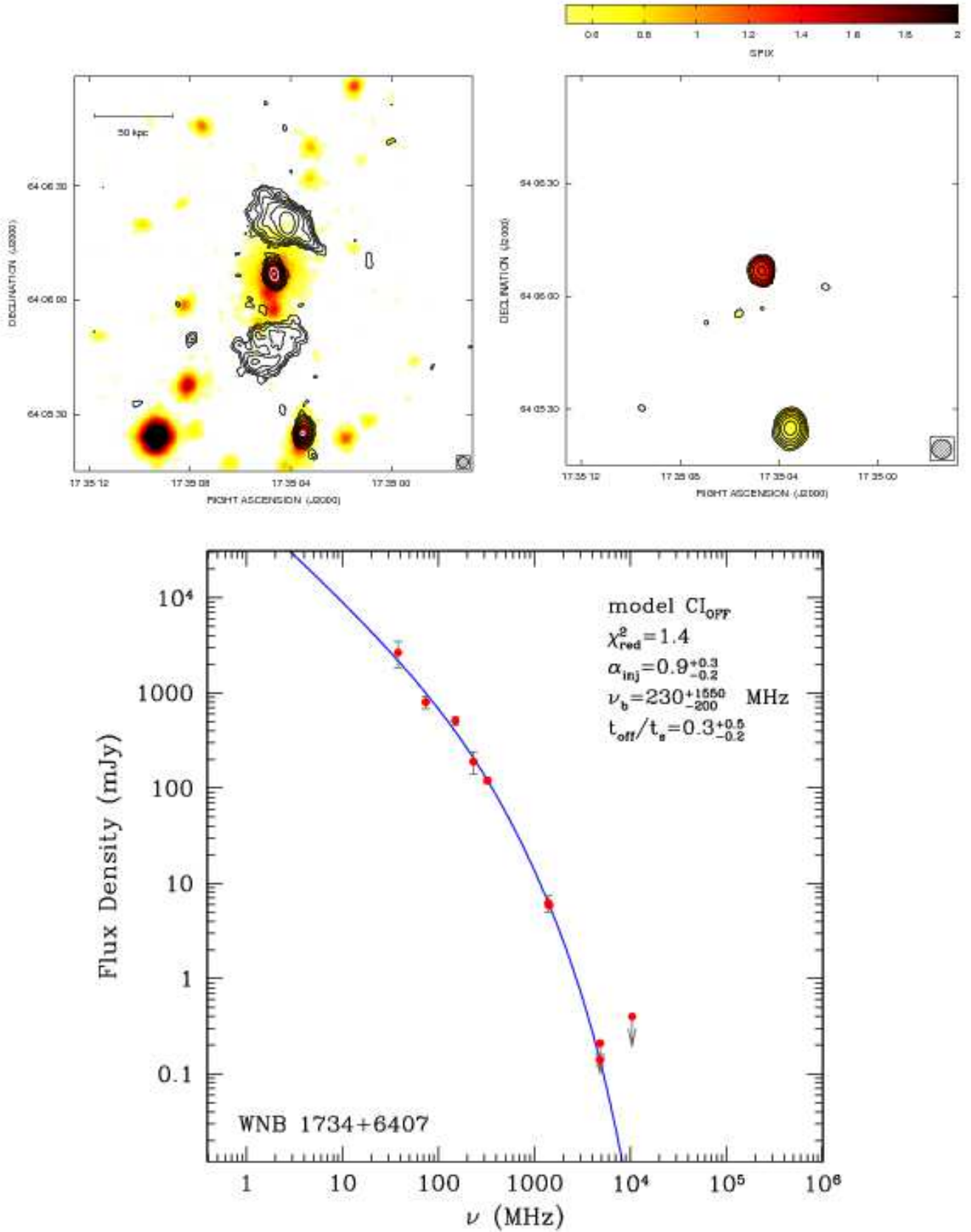}

\caption{WNB 1734+6407. Top-left panel image has been taken at 1.4\,GHz with the VLA in A+B+C configuration, the restoring beam FWHM is $2.9\arcsec\times 2.8\arcsec$. The contours of the radio intensity are overlaid on the optical DSS2 image. The top-right panel show the overlay of the spectral index image between 1.4 and 4.8\,GHz with the 4.8\,GHz contour levels of the VLA  C+D array image. The restoring beam FWHM is $5.2\arcsec\times 5.1\arcsec$. Both the 1.4 and 4.8\,GHz images have the same sensitivity level of $10\,\mu$Jy/beam. Radio contours start from a level of $3\sigma$-rms and scale by $\sqrt{2}$. In the bottom panel we show the integrated spectrum of the source along with the best fit of the synchrotron model.}
\end{center}

\label{fig1}
\end{figure*}

\begin{figure*}
\begin{center}
\includegraphics[width=18cm]{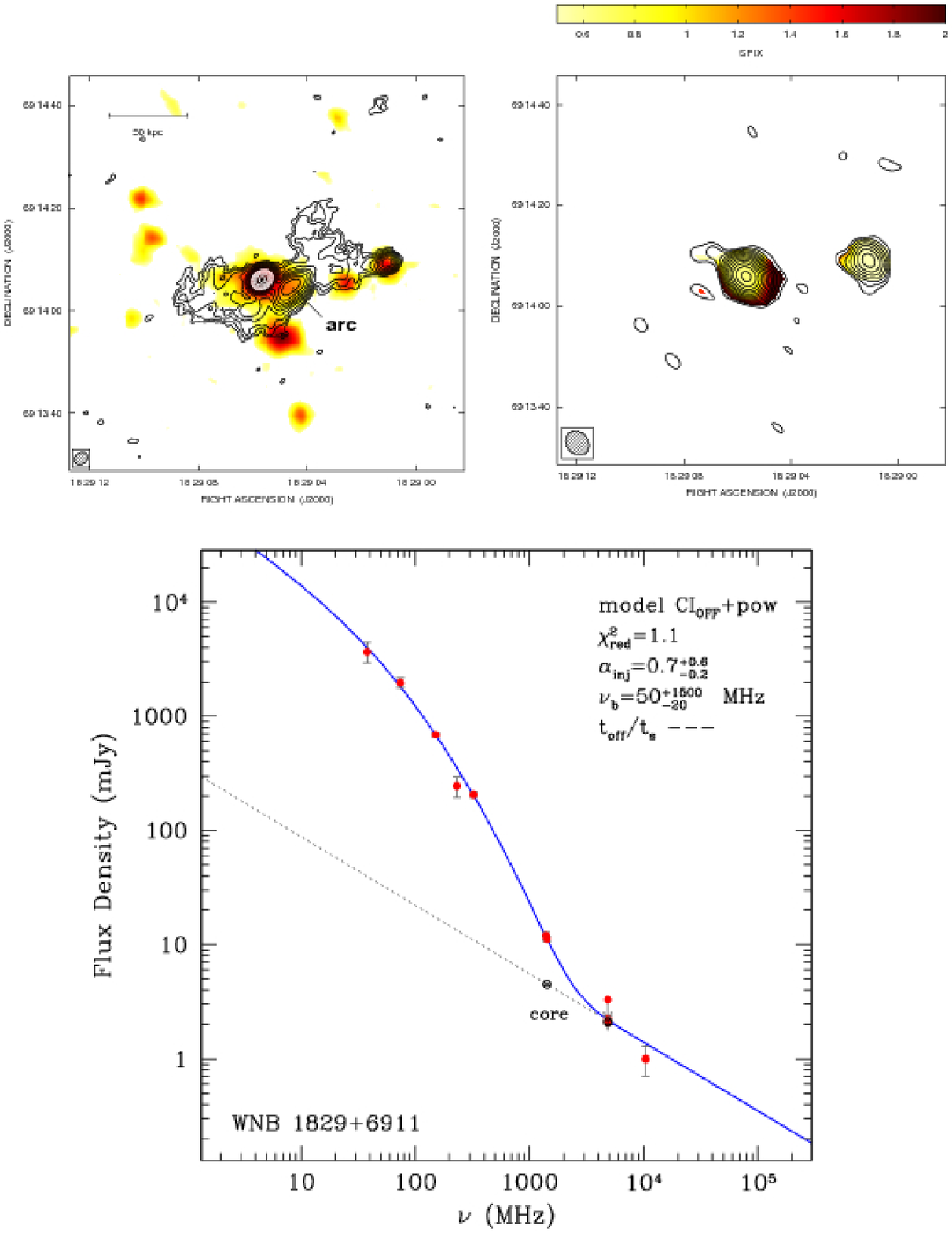}
\caption{WNB 1829+6911. Top-left panel image has been taken at 1.4\,GHz with the VLA in A+B+C configuration, the restoring beam FWHM is $2.9\arcsec\times 2.3\arcsec$. The contours of the radio intensity are overlaid on the optical DSS2 image.
The top-right panel show the overlay of the spectral index image between 1.4 and 4.8\,GHz with the 4.8\,GHz contour levels of the VLA C+D array image. The restoring beam FWHM is $5.2\arcsec\times 4.3\arcsec$. Both the 1.4 and 4.8\,GHz images have the same sensitivity level of $13\,\mu$Jy/beam. Radio contours start from a level of $3\sigma$-rms and scale by $\sqrt{2}$. In the bottom panel we show the integrated spectrum of the source along with the best fit of the synchrotron model.}
\end{center}
\label{fig2}
\end{figure*}

Three of these 200 sources show  a quasi-exponential drop of their integrated 
spectrum with a spectral index as steep as $\alpha > 1.5$ at 1.4\,GHz. 
These sources are WNB1734+6407, WNB1829+6911, and WNB1851+5707. 

The spectral curvatures, calculated between 
151 and 325 MHz (for $\alpha_{\rm low}$) and between 1400 and 4850 MHz (for $\alpha_{\rm high}$), for the three sources 
are respectively $SCP=1.1, -0.4,$ and $1.1$. Thus, WNB1734+6407 and WNB1851+5707 can be
 classified as dying sources candidates, while WNB1829+6911 would be a restarting source candidate. 

All the three radio sources are
associated with elliptical galaxies (see  Tab.\,1 for a list of their basic properties).
As we will show in the next section, it turn out that 
WNB1851+5707 is indeed composed by two distinct dying galaxies that are blended together in the WENSS.
The two redshifts listed in  Tab.\,1 have been obtained with the DOLORES spectrograph
installed at the 3.5-m Galileo telescope on the Roque
de los Muchachos in La Palma, Spain (see Parma et al. 2007 for more details on the optical photometry).  

\begin{figure*}
\begin{center}
\includegraphics[width=18cm]{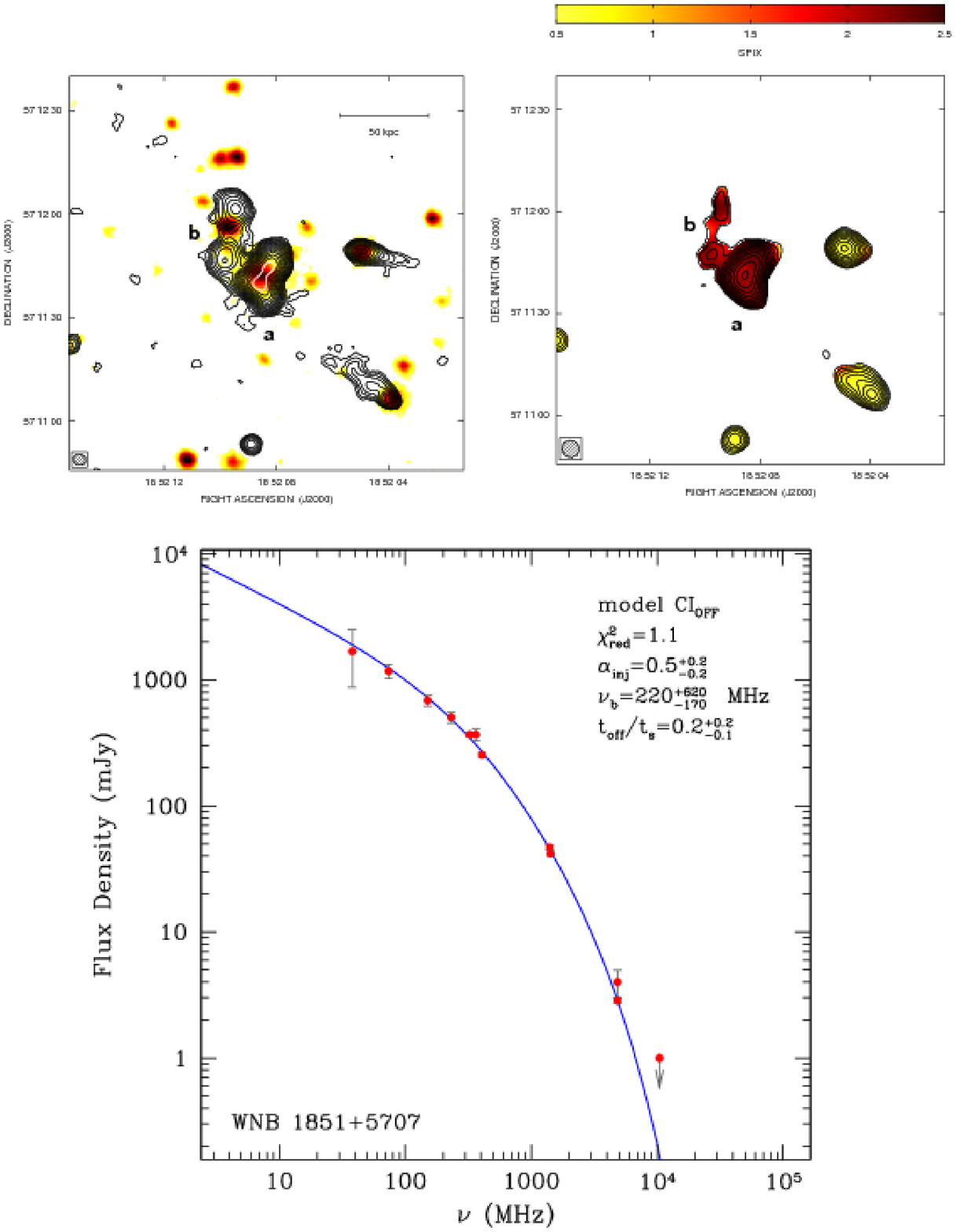}
\caption{WNB 1851+5707. Top-left panel image has been taken at 1.4\,GHz with the VLA in A+B+C configuration, the restoring beam FWHM is $3.7\arcsec\times 3.3\arcsec$. The contours of the radio intensity are overlaid on the optical DSS2 image. WNB 1851+5707 is composed by two distinct dying
 sources labeled $a$ and $b$.
The top-right panel show the overlay of the spectral index image between 1.4 and 4.8\,GHz with the 4.8\,GHz contour levels of the VLA C+D array image. The restoring beam FWHM is $5.3\arcsec\times 5.0\arcsec$. The sensitivity levels of the 1.4 and 4.8\,GHz images are 17 and $9\,\mu$Jy/beam, respectively. Radio contours start from a level of $3\sigma$-rms and scale by $\sqrt{2}$. In the bottom panel we show the integrated spectrum of the source along with the best fit of the synchrotron model. The radio spectrum is representative of WNB 1851+5707a, the brighter of the two dying sources (see text).}
\end{center}
\label{fig3}
\end{figure*}

In order to determine whether these sources were really dying objects or relic lobes associated with active radio galaxies, we observed their continuum emission in details with the VLA. The brighter and more compact of the three, WNB1851+5707, has been also observed with the MERLIN.

\begin{figure*}
\begin{center}
\includegraphics[width=18cm]{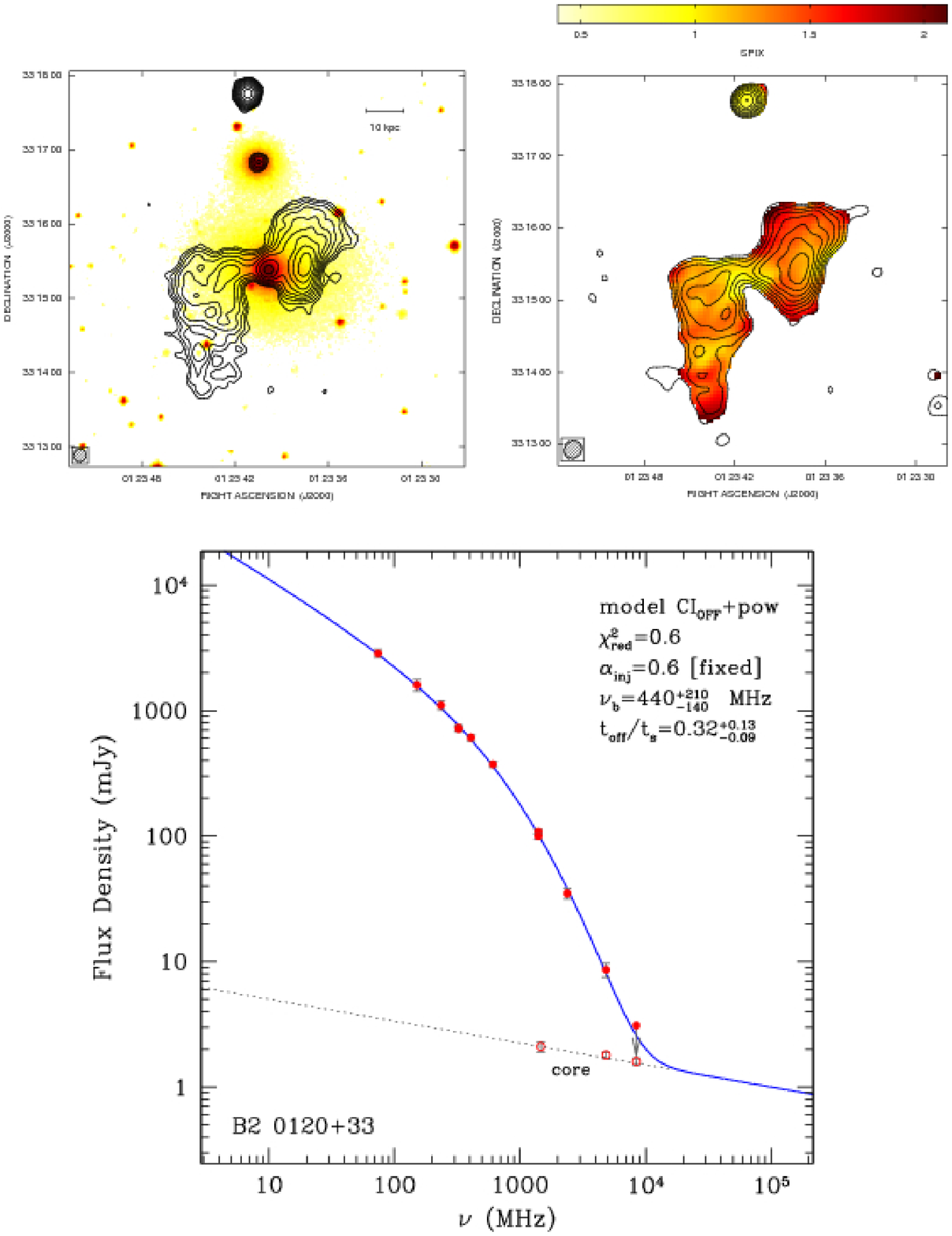}
\caption{B2 0120+33. Top-left panel image has been taken at 1.4\,GHz with the VLA in B+C configuration, the restoring beam FWHM is $11.2\arcsec \times 10.4\arcsec$. The contours of the radio intensity are overlaid on the optical DSS2 image. The top-right panel show the overlay of the spectral index image between 324\,MHz and 1.4\,GHz with the 1.4\,GHz contour levels of the VLA C array image. The restoring beam FWHM is $16\arcsec\times 14.5\arcsec$. The sensitivity levels of 324\,MHz and 1.4\,GHz images are 1.4 mJy/beam and $90\,\mu$Jy/beam, respectively. Radio contours start from a level of $3\sigma$-rms and scale by $\sqrt{2}$. In the bottom panel we show the integrated spectrum of the source along with the best fit of the synchrotron model.}
\end{center}
\label{fig4}
\end{figure*}

\begin{figure*}
\begin{center}
\includegraphics[width=18cm]{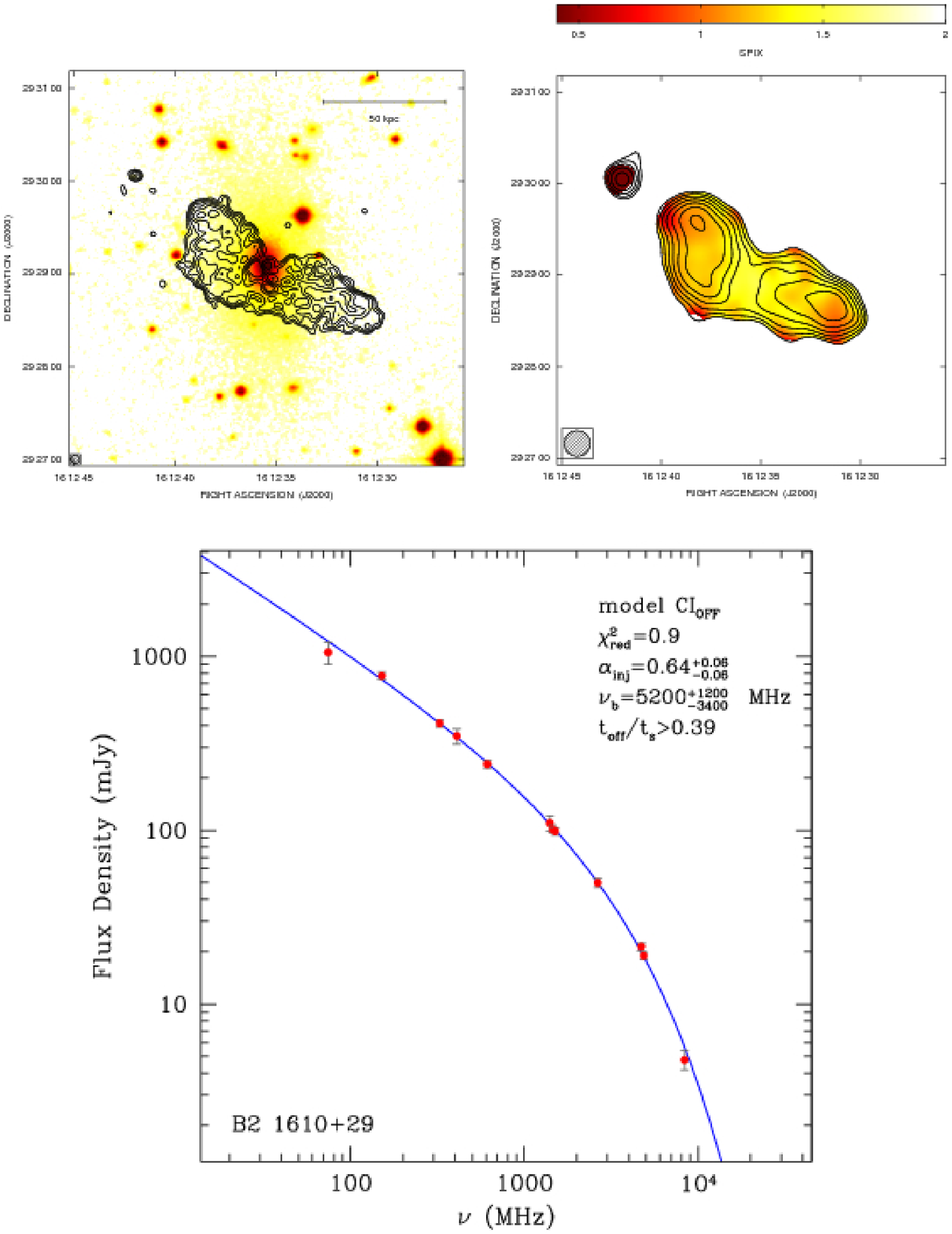}
\caption{B2 1610+29. Top-left panel image has been taken at 1.4\,GHz with the VLA in B+C configuration, the restoring beam FWHM is $5.5\arcsec\times 5.5\arcsec$. 
The contours of the radio intensity are overlaid on the optical DSS2 image. The top-right panel show the overlay of the spectral index image between 1.4 and 4.7\,GHz with the 4.7\,GHz contour levels of the VLA D array image. The restoring beam FWHM is $17\arcsec\times 17\arcsec$. The sensitivity levels of the 1.4 and 4.7\,GHz images are 210 and $80\,\mu$Jy/beam, respectively. Radio contours start from a level of $3\sigma$-rms and scale by $\sqrt{2}$. In the bottom panel we show the integrated spectrum of the source along with the best fit of the synchrotron model.}
\end{center}
\label{fig5}
\end{figure*}

\begin{figure*}
\includegraphics[width=18cm]{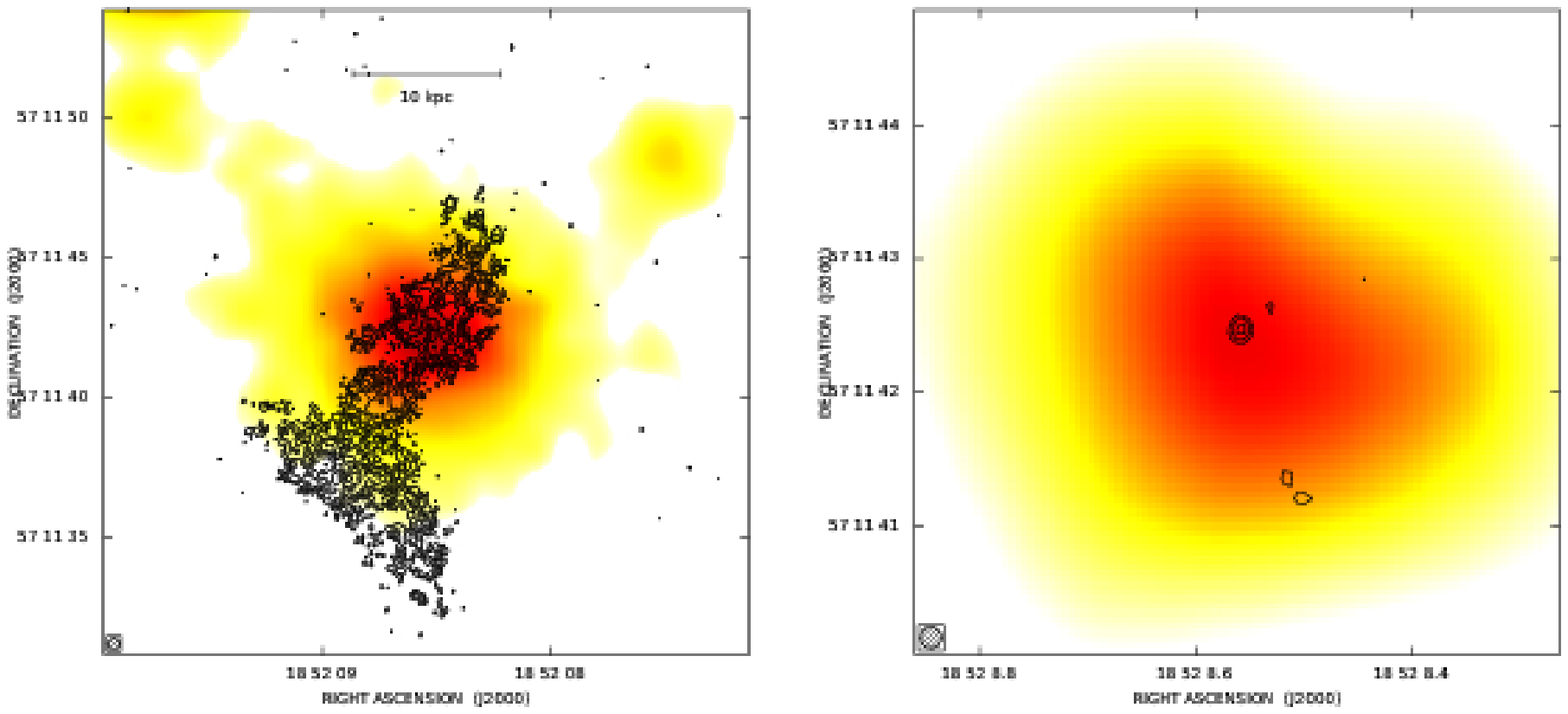}
\caption{MERLIN images WNB 1851+5707a at 1.4 GHz. Left, MERLIN+VLA radio contours at $0\farcs4$ resolution overlaid the DSS2 image. Contours start at 0.1 mJy/beam and increase by a factor of $\sqrt{2}$. Right, zoom of the galaxy core. The MERLIN contours at 0\farcs16 resolution are overlaid to the optical DSS2 image.  Contours start at 0.012 mJy/beam and increase by a factor of $\sqrt{2}$.  WNB 1851+5707b is not visible in the field of view of the MERLIN observation.}
\label{MERLIN}
\end{figure*}

\subsection{The B2 sample}
The B2 bright sample described in Colla et al. (1975) has been selected with very 
similar criteria to the WENSS minisurvey: radio sources with a flux density $>200$ mJy at 408 MHz 
have been identified with bright nearby galaxies. This complete sample counts 54 elliptical galaxies and thus we may expect it may contain 
a few dying sources. For the B2 sample detailed VLA images are available. We searched for sources with relaxed radio morphologies and extreme spectral curvatures 
 mostly calculated between 151 and 408 MHz (for $\alpha_{\rm low}$) and between 1400 and 4850 MHz (for $\alpha_{\rm high}$).
Indeed, the prototypical dying source B2 0924+30 with $SPC=0.8$ and the restarting source\footnote{Note that in this restarting source the active core-jets components account for just a small fraction of the source's flux density, which is dominated by the steep spectrum lobes in the considered frequency window.} 3C338 (alias B2 1626+39) with $SPC=0.7$ belong
 to this sample. 

Besides B2 0924+30 and B2 1626+39, we searched for other dying or re-starting sources candidates.
We found that the radio sources B2 0120+33 and B2 1610+29, with a 
spectral curvature of respectively $SPC=1.5$ and $SPC=0.8$, can be indeed considered dying radio source candidates. 
The radio source B2 0120+33 is associated to the elliptical galaxy NGC 507 at the center of the Zwicky Cluster 0107.5+3212. The
source has a weak radio core and the large scale radio emission appears very faint and diffuse at a resolution of few arcseconds 
 even at 1.4 GHz (Parma et al. 1986).   
The second candidate, B2 1610+29, is associated to the nearby cD central galaxy NGC 6086 in Abell 2162 and
 has relaxed morphology in the VLA 1.4 GHz image of Parma et al. (1986).  More recently the source has been imaged with the 
GMRT at 610 MHz by Giacintucci et al. (2007), who found a very similar morphology.

By using all the useful data that are available in 
the VLA archive since then, we studied in detail the spectral properties of these objects in order to determine if they really are 
dying sources or not.

\section{VLA observations and data reduction}

\subsection{New observations}
We observed the three dying radio galaxy candidates from the WENSS minisurvey at 1.4 and 4.8 GHz with VLA in 
various configurations. A summary of this campaign of observations, including 
the VLA configuration, frequency date and length of observations, is reported in Tab.~\ref{obssum}.
A bandwidth of 50 MHz was used for each of the two frequencies.

Calibration and imaging were performed with the Astronomical Image Processing System (AIPS), following the standard procedure:
Fourier-Transform, Clean and Restore. Self-calibration was applied to remove residual
phase variations. Data from different arrays were combined to 
improve uv-coverage and sensitivity. We combined the A, B and C arrays at 1.4\,GHz and the C and D at 4.8\,GHz. 
Each combined data set was self-calibrated. The relevant image parameters are reported in Tab.~\ref{imapar}.

In order to estimate the flux density and the spectral index of the three sources 
the primary beam correction was applied to all the images by applying the AIPS task PBCOR. For the purposes of 
the spectral index imaging, we convolved the 1.4\,GHz images to the slightly larger
beam of the 4.8\,GHz ones. However, we emphasize that the angular resolution of the final images at 
the two frequencies were already well matched, not just by filtering the data, but by having comparable
intrinsic coverage over the relevant spatial frequencies. Furthermore, both data sets have nearly the same 
sensitivity levels.

\begin{table}[h]
\caption[]{VLA observations summary.}
\begin{center}
\begin{tabular}{ccccc}
\hline
\noalign{\smallskip}
   Source& Array & Frequency & Date &  Duration\\
         &     & (GHz)     &  &          (hours)\\
\noalign{\smallskip}
\hline
\noalign{\smallskip}
WNB1734+6407 &  A & 1.4& 21-Apr-2002 & 5.2\\
             &  B & 1.4& 17-Jul-2002 & 2.7\\
             &  C & 1.4& 20-Dec-2002 & 2.4\\
             &  C & 4.8& 21-Dec-2002 & 6.2\\
             &  D & 4.8 & 07-Dec-2001 & 3.0\\
\hline
WNB1829+6911 &  A & 1.4& 11-Apr-2002 & 2.6\\
             &  B & 1.4& 17-Jul-2002 & 2.7\\
             &  C & 1.4& 20-Dec-2002 & 2.4\\
             &  C & 4.8& 21-Dec-2002 & 0.4\\
             &  D & 4.8 & 07-Dec-2001 & 3.1\\
\hline
WNB1851+5707 &  A  & 1.4& 26-Mar-2002 & 3.0\\
             &  B & 1.4& 17-Jul-2002 & 0.8\\
             &  C & 1.4& 22-Jun-2002 & 2.8\\
             &  C & 4.8& 15-Dec-2002 & 4.0\\
             &  D & 4.8 & 08-Dec-2001 & 3.2\\
\noalign{\smallskip}
\hline
\end{tabular}
\end{center}
\label{obssum}
\end{table}

The VLA total intensity iso-contours of WNB 1734+6407, WNB 1829+6911, and WNB 1851+570 at 1.4 and 4.8\,GHz
are respectively shown in top left and right panels of Figs.\,1 to 3. The 1.4\,GHz contours are overlaid to
the optical image of the red Digital Sky Survey (DSS2), while the 4.8\,GHz contours are overlaid to the spectral index image between
the two radio frequencies.

\subsection{Archival data}
We recover from the VLA archive all the useful data that are available for the radio sources B2 0120+33 and B2 1610+29. A summary of relevant parameters of the images produced is listed in Tab.\,3. 

The source B2 0120+33 has a useful 325\,MHz B-array observation in the VLA archive (project AC0845). 
The observation was acquired in line mode with 15 spectral channels in a bandwidth of 6 MHz per each of the two IFs. The source 
 3C 48 was used as amplitude, phase, and bandpass calibrator.
In the final imaging the data were averaged to 6 channels and mapped using a wide-field imaging technique, 
which corrects for distortions in the image caused by the non-coplanarity of the VLA over a wide field of view. A set of 
overlapping fields was used to cover an area of about $2.5\degr\times2.5\degr$ around the radio source. 
All these fields were included in CLEAN and used for several loops of phase self-calibration. The central frequency of the final images is 324 MHz. At 1.4\,GHz we produced a B+C image at a resolution of about 10\arcsec. This image is 
shown in top-left panel of Fig.\,4. The resolution of the C array 1.4\,GHz image alone well matches that of the 325\,MHz image and the two have been used to derive the spectral index image shown in top-right panel of Fig.\,4. It is worthwhile to mention that flux densities of both the  325\,MHz and the
 1.4\,GHz images are perfectly consistent with that of the WENSS and the NVSS. At 4.8\,GHz we were able to combine the B and C array. Only the core 
and the brightest part of the lobes are detected in this image that can be used however to derive an integrated flux density for the radio source. 
At 8.6\,GHz we recover a D array image in which only the core of the radio galaxy is visible. 

The source B2 1610+29 has an angular size of 
about two arcmin and has a very relaxed morphology which is completely resolved out by the A configuration of the VLA at 1.4 GHz.
Thus, we opted to combine just the B and C array at this frequency, obtaining a final resolution of $5\farcs5 \times 5\farcs5$ and a
rms noise level of 70 $\mu$Jy/beam. This is the highest resolution image we present for this source and it is shown as iso-contours 
in top-left panel of Fig.\,5. The flux density of our B+C image is consistent, within the errors, with the NVSS flux. Therefore, we are
confident to have recovered most of flux density from the extended source structure. We also produced images at 1.4 and 4.7\,GHz in C and 
D configuration, respectively. These images, which have been restored with the same resolution of $17\arcsec \times 17\arcsec$ and corrected 
from the primary beam attenuation, are characterized by roughly the same uv-coverage and have been used to produce the spectral index
image shown in the top-right panel of Fig.\,5. Finally, for the purposes of the analysis of the source integrated spectrum, we reduced separately
the two IFs of the L band observation obtaining a measure of the source flux densities at 1452 and 1502\,MHz.

\begin{table}[h]
\caption[]{Image parameters summary.}
\begin{center}
\begin{tabular}{ccccc}
\hline
\noalign{\smallskip}
   Source& Array & $\nu$ & Beam      & $\sigma$   \\
         &       & (GHz)       & (arcsec)    & ($\mu$Jy/beam)  \\
\noalign{\smallskip}
\hline
\noalign{\smallskip}
WNB1734+6407&  A+B+C &  1.4    & $2.9 \times 2. 8$ & 10   \\
  &C+D &  4.8    & $5.2 \times 5.1$ & 10     \\
\hline
 WNB1829+6911&  A+B+C &  1.4     & $2.9 \times 2.3$ & 13  \\
  &C+D &  4.8    & $5.2 \times 4.3$ & 13  \\
\hline
 WNB1851+5707&  A+B+C &  1.4    & $3.7 \times 3.3$ & 17  \\
 & C+D &  4.8     & $5.3 \times 5.0$ & 9  \\
\hline
 B2 0120+33 &  B   &  0.324  & $16.0 \times 14.5$ & 1400  \\
            &  C   &  1.4    & $13.4 \times 12.5$ & 50  \\
            &  B+C &  1.4    & $11.2 \times 10.4$ & 80  \\
            &  B+C &  4.8    & $4.0 \times 4.0$ & 30  \\
            &  D   &  8.4    & $8.3 \times 5.1$ & 70  \\

\hline
 B2 1610+296&  B+C &  1.4    & $5.5 \times 5.5$ & 70  \\
            &  C   &  1.4    & $17.0 \times 17.0$ & 210  \\
            &  D   &  4.7    & $17.0 \times 17.0$ & 80  \\
\noalign{\smallskip}
\hline

\end{tabular}
\end{center}
\label{imapar}
\end{table}

\subsection{Comments on individual sources}

The VLA integrated flux density and spectral index for the individual source components are listed in Tab.\,4.

\subsubsection{WNB 1734+6407}
Characterized by two relaxed lobes lacking hot-spots, the radio morphology of WNB1734+6407
resembles that of B2 0924+30, which is considered the prototype of fossil radio galaxies (Cordey 1987, but
 see also Jamrozy et al. 2004 for more recent data on this source).
The radio spectrum of the fossil lobes of WNB 1734+6407 is so steep that their surface brightness at
6 cm is below the sensitivity level of our observations. Thus, we can only place lower limits on
the spectral index which result in $\alpha>2.7$ and  $\alpha>2.3$ for the northern and southern lobe,
respectively. We also observe a slightly extended component coincident with the galaxy center. Also this
feature has a quite steep spectral index, $\alpha=1.3$, and its nature remains unclear. In the VLA A-array
 image (not shown) this feature is elongated in N-S direction with an extent of about 7 kpc $\times$ 14 kpc, resembling 
a core-jet morphology.

\begin{figure*}
\begin{center}
\includegraphics[width=16cm]{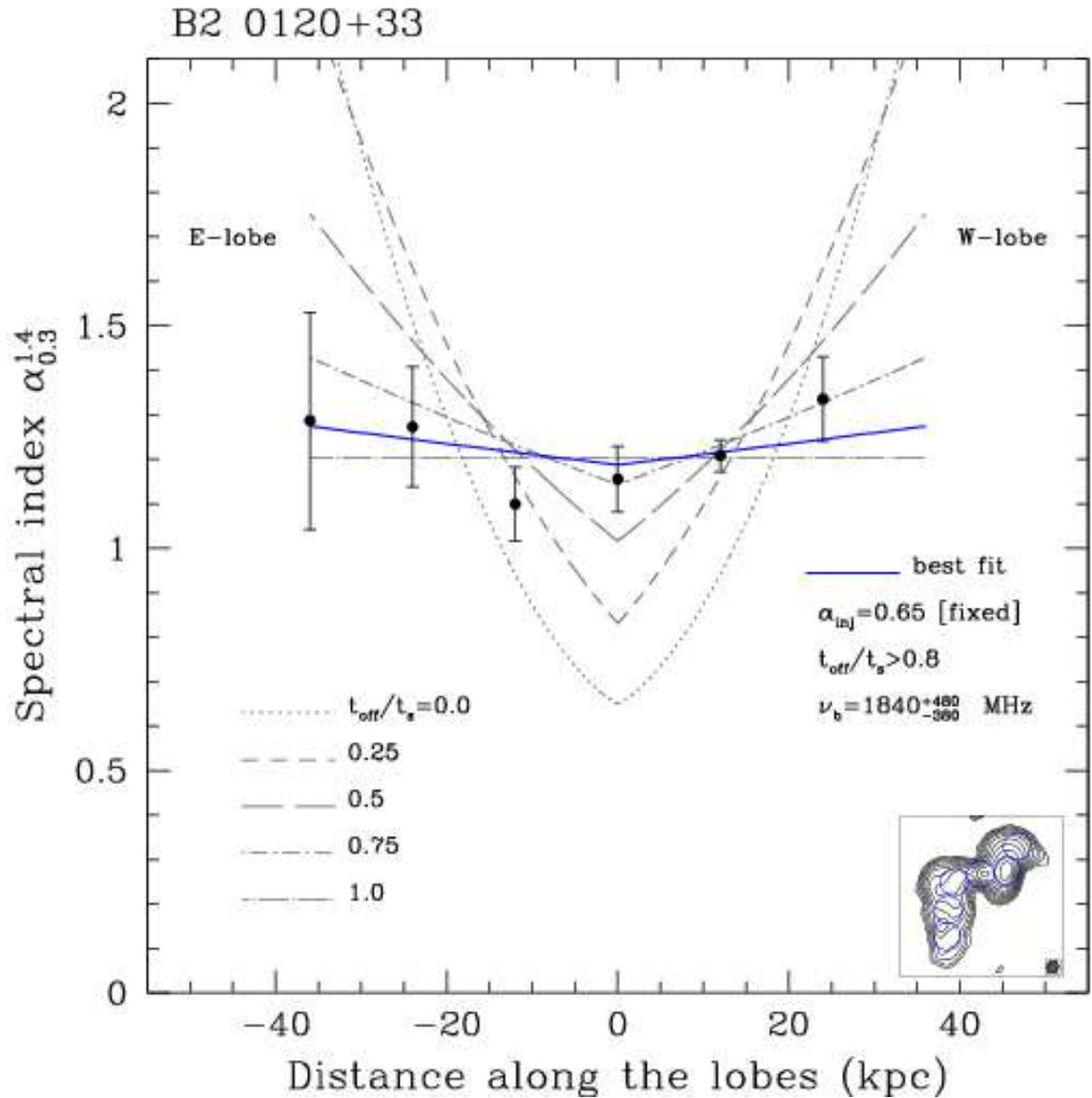}
\caption{B2 0120+33 spectral index profile. Each point represents the average over a circular region as shown in the inset. The solid line is the
best fit of the $CI_{\rm OFF}$ model described in the text. The reference dashed lines correspond to different relative durations of the dying phase.}
\end{center}
\label{fig6}
\end{figure*}

\begin{figure*}
\begin{center}
\includegraphics[width=16cm]{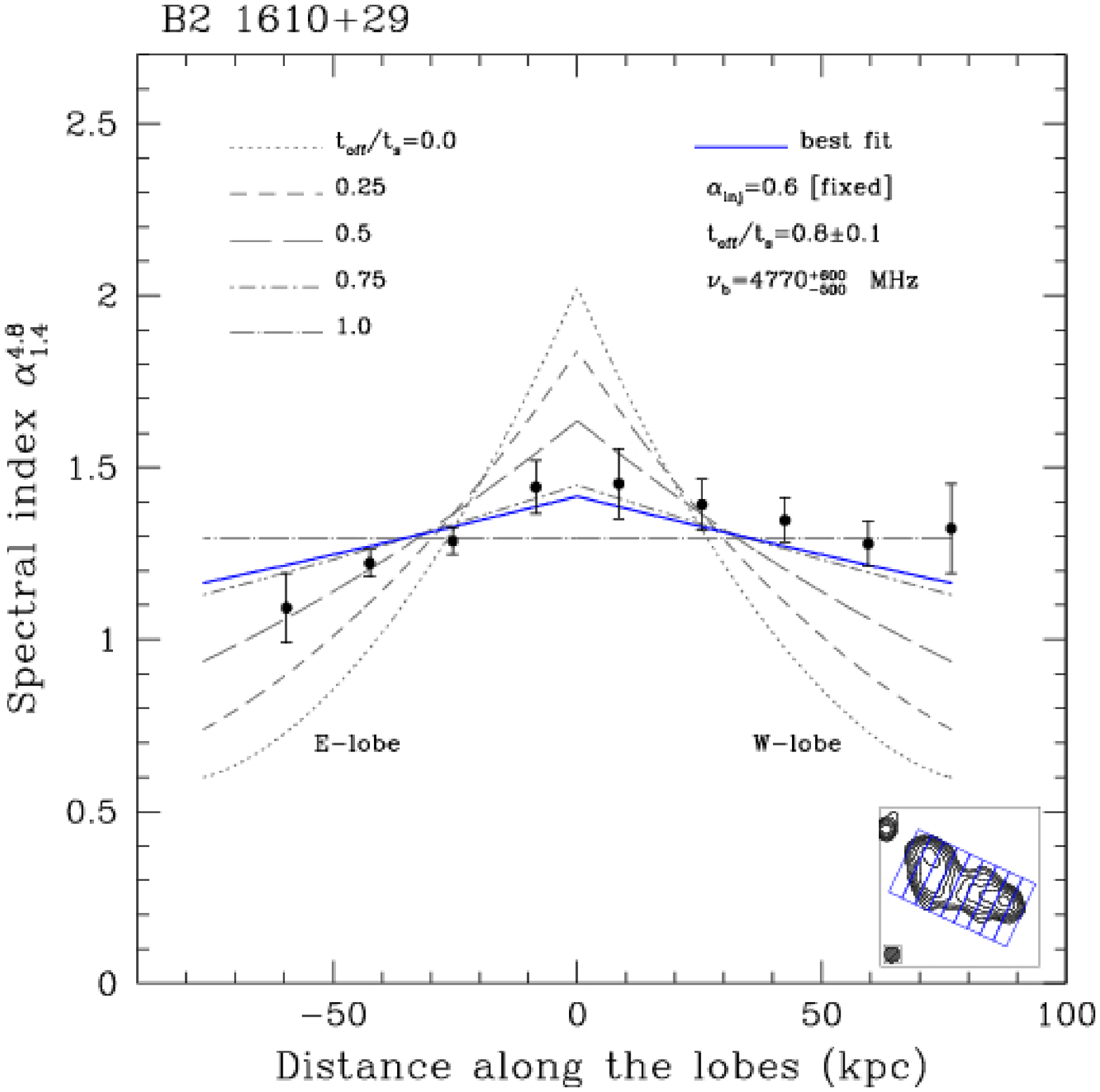}
\caption{B2 1610+29 spectral index profile. Each point represents the average over a rectangular region as shown in the inset. The solid line is the
best fit of the $CI_{\rm OFF}$ model described in the text. The reference dashed lines correspond to different relative durations of the dying phase.}
\end{center}
\label{fig4}
\end{figure*}

\subsubsection{WNB 1829+6911}
The radio appearance of WNB1829+6911 is virtually identical to that of 3C 338, a
nearby radio source associated with the central dominant galaxy in the cooling flow
cluster Abell 2199  (Jones \& Preston 2001). In both sources we observe the presence of fossil plasma remaining
from a previous activity in conjunction with a restarting core.  The extended emission of
WNB1829+6911 has a very steep spectrum, $\alpha > 1.6$ and $\alpha > 1.9$ respectively for the W and E-lobe. 
The source exhibits a bright core with a much flatter spectrum, $\alpha = 0.6$. Most probably this region of the 
radio source is powered by a new couple of restarting jets. The core emission is responsible for the high-frequency flattening
seen in the source integrated spectrum (Fig.\,2 bottom panel). An arc-like feature with a very steep
 spectrum ($\alpha > 2.5$) can be seen on the SW side of the core (Fig.\,2 top-left panel).

\begin{table*}
\caption[]{VLA integrated flux densities and spectral indexes of sources components.}
\label{fluxes}
\begin{center}
\begin{tabular}{ccccccc}
\hline
\noalign{\smallskip}
   Source& Component      & size            &$S_{\rm 1.4}$   & $S_{\rm 4.8}$     &  $\alpha_{1.4}^{4.8}$   \\
         &                & kpc$\times$kpc  & mJy            &   mJy             &                        \\
\noalign{\smallskip}
\hline
\noalign{\smallskip}
   WNB1734+6407 &  N-Lobe    &  $63\times 42$   & $2.9\pm0.03$  &     $<0.1$        & $>2.7$\\
                &  S-Lobe    &  $60\times 42$   & $1.7\pm0.03$  &     $<0.1$        & $>2.3$ \\
                &  Core-Jet? &  $14\times 7$    & $0.7\pm0.01$  &     $0.14\pm0.02$ & $1.3\pm0.1$\\
\noalign{\smallskip}
                &  Total   &  $120\times 62$  & $5.9\pm0.1$   &     $0.14\pm0.02$ & $3.0\pm0.1$\\

\noalign{\smallskip}
\hline
\noalign{\smallskip}
   WNB1829+6911 &  E-Lobe & $72\times 47$  & $2.1\pm0.07$  & $<0.2$        & $>1.9$ \\
                &  W-Lobe & $69\times 41$  & $1.4\pm0.06$  & $<0.2$        & $>1.6$ \\
                &  Arc    & $33\times 16$  & $2.1\pm0.05$  & $<0.1$        & $>2.5$ \\
                &  Core   & --             & $4.5\pm0.02$  & $2.1\pm0.03$  & $0.62\pm 0.01$ \\
\noalign{\smallskip}
                &  Total  & $135\times 61$ & $11.2\pm0.1$  & $2.1\pm0.03$  & $1.4\pm0.01$\\

\noalign{\smallskip}
\hline
\noalign{\smallskip}
   WNB1851+5707a &  Total  & $15\times 30$  & $41.8\pm0.08$   & $2.8\pm0.04$   & $2.2\pm0.1$  \\
   
\noalign{\smallskip}
   WNB1851+5707b &  N-Lobe & $23\times 17$  & $2.6\pm0.07$   & $0.17\pm0.02$   & $2.2\pm0.3$  \\
                 &  S-Lobe & $19\times 19$  & $2.3\pm0.05$   & $0.13\pm0.02$   & $2.3\pm0.3$  \\
\noalign{\smallskip}
                 &  Total  & $50\times 25$  & $4.9\pm0.06$   & $0.3\pm0.03$    & $2.3\pm0.2$  \\

\noalign{\smallskip}
\hline
\noalign{\smallskip}
   B2 0120+33   &  E-Lobe & $47\times 24$  & $43.6\pm0.3$  & $3.64\pm0.6$  & $2.0\pm0.14$ \\
                &  W-Lobe & $32\times 22$  & $56.5\pm0.2$ & $4.5\pm0.5$  & $2.1\pm0.1$ \\
                &  Core   & --             & $2.1\pm0.2$  & $1.8\pm0.1$   & $0.13\pm 0.09$ \\
\noalign{\smallskip}
                &  Total  & $73\times 40$  & $99.5\pm1.5$  & $8.6\pm1.1$  & $2.0\pm0.1$\\

\noalign{\smallskip}
\hline
\noalign{\smallskip}
   B2 1610+29    &  Total   &  $100\times 30$  & $100\pm1.4$ &     $15.5\pm0.3$ & $1.5\pm0.02$\\

\noalign{\smallskip}
\hline
\end{tabular}
\end{center}
\label{phypar}
\end{table*}

\subsubsection{WNB 1851+5707}
Our VLA images reveal that WNB1851+5707 is in reality composed by two distinct dying radio
galaxies. For source WNB1851+5707a  we measured a spectral index $\alpha=2.2$ while  WNB1851+5707b we measure $\alpha=2.3$. 
This association is a really
intriguing fact considering the rarity of this kind of sources. 
According to the interpretation that these are relic sources, we found that,
at the sensitivity limits of our VLA observations, they possess neither radio jets 
on kpc scale nor bright hot spots inside the fossil lobes.
The two hosting galaxy have nearly the same redshifts (see Tab.\,1), making the possibility of a 
spatial casual coincidence even more unlike.

It should be noted that WNB1851+5707 is located at the center of a galaxy cluster, see Sect.\,6. Thus, it could be that the 
two dying lobes of WNB1851+5707b are indeed two previous outbursts 
of WNB1851+5707a that are emerging from the cluster center because of the buoyancy forces.  
Although this scenario seems less likely, in view of the almost perfect spatial coincidence of WNB1851+5707b with the second
 galaxy. 

\subsubsection{B2 0120+33}
The radio source is asymmetric, the west lobe being brighter and smaller that the east one. The radio 
emission is very weak and diffuse. A similar morphology has been observed in GMRT images at 235 and 610 MHz by Giacintucci et al. 2010 (in preparation).
At arcsecond resolution the only compact structure is the faint 
unresolved core. There is no evidence of kpc-scale jets in any of our images. Hence, the extended lobes of 
 B2 0120+33 have been produced by an earlier phase of activity. 
 
\subsubsection{B2 1610+29}
The morphology of B2 1610+29 is exactly the one we would expect for a dying radio source: two symmetric 
 and relaxed radio lobes sit on the opposite side of the host galaxy, there are no jets nor hot-spots in 
the lobes, and the host galaxy lacks a bright radio core. All these features are strong indications that B2 1610+29 
 is a dying source where only the fading lobes are still visible. The source has an overall linear size of about 100 kpc
 with a small offset of the source major axis to south-est with respect to the position of the galaxy. The host galaxy of B2 1610+29  
has been recently studied with the VLBA at 5\,GHz by Liuzzo et al. (2010).  They did not detected any compact radio structure 
in their high resolution images. Also this results suggests that most likely B2 1610+29 is a dying radio source.

\begin{table*}
\caption[]{Integrated spectra.}
\begin{center}
\begin{tabular}{lcccccc}
\hline
   Source& Frequency & $S_{\nu}^{\rm catalogued}$ & $f_{Baars}$& Reference & $S_{\nu}^{\rm confusion}$ & $S_{\nu}$ \\
         &      MHz  &   mJy    & & & mJy & mJy  \\
\hline
\noalign{\smallskip}
   WNB 1734+6407&  38   & $2699 \pm 818$  &0.818 & 8C, Hales et al. 1995& $50\pm 12$     & $ 2649\pm 818$ \\
                &  74   & $ 810 \pm 120$  &1.0   & VLSS, Cohen et al. 2007& $ 8\pm 3$  & $ 802 \pm 120$ \\
                &  151  & $ 524 \pm 52$   &0.904 & 6CIII, Hales et al. 1990& $ 14\pm 2 $ & $ 510 \pm  52$ \\
                &  232  & $ 200 \pm 50$   &1.0   & MIYUN, Zhang et al. 1997& $ 9.8\pm 1.1$  & $ 190 \pm 50$ \\
                &  325  & $ 123 \pm 7.3$  &1.0   & WENSS, Rengelink et al. 1997& $ 2.8\pm 0.5 $  & $ 120 \pm 7$ \\
                &  1400 & $ 7.2 \pm 1.2$  &1.0   & NVSS, Condon et al. 1998& $ 1.04\pm 0.04$  & $ 6.2 \pm 1.2$ \\
                &  1425 & $ 5.9 \pm 0.1$  &1.0   & Effelsberg 100-m, this work& $ - $           & $ 5.9 \pm 0.1$ \\
                &  4850 & $ 0.2 \pm 0.02$ &1.0   & VLA, this work& $ 0.63\pm 0.07$ & $ <0.21 $ \\
                &  4850 & $ 0.14\pm 0.02$ &1.0   & Effelsberg 100-m, this work& $ - $           & $ 0.14\pm 0.02$ \\
                & 10450 & $<0.4$          &1.0   & VLA, this work& $ - $           & $ <0.4 $ \\
   \hline
\noalign{\smallskip}
   WNB 1829+6911&  38   & $ 3681 \pm 740$  &0.818  &  8C, Hales et al. 1995& $ 21\pm 27$       & $ 3661 \pm 740$ \\
                &  74   & $ 1960 \pm 200$  &1.0    &  VLSS, Cohen et al. 2007& $ 2.0\pm 0.76$    & $ 1958 \pm 200$ \\
                &  151  & $ 694 \pm 31$    &0.904  &  6CIV, Hales et al. 1991& $ 7.2\pm 4.9$     & $ 687  \pm 31$ \\
                &  232  & $ 250 \pm 50$    &1.0    &  MIYUN, Zhang et al. 1997& $ 5.2\pm2.7 $     & $ 245 \pm 50$ \\
                &  325  & $ 207 \pm 9$     &1.0    &  WENSS, Rengelink et al. 1997& $ 1.2\pm 0.22$    & $ 209 \pm 9$ \\
                &  1400 & $ 12.6 \pm 0.9$  &1.0    &  NVSS, Condon et al. 1998& $ 0.71\pm 0.02$   & $ 11.9 \pm 0.9$ \\
                &  1425 & $ 11.2 \pm 0.1$  &1.0    &  VLA, this work& $ - $             & $ 11.2 \pm 0.1$ \\
                &  4850 & $ 2.25 \pm 0.12$ &1.0    &  CATS, Verkhodanov et al. 1997& $ - $             & $ 2.25 \pm 0.12$ \\
                &  4850 & $ <3.3       $   &1.0    &  Effelsberg 100-m, this work& $ - $             & $ <3.3$ \\
                &  4860 & $ 2.1 \pm 0.03$  &1.0    &  VLA, this work& $ - $             & $ 2.1\pm 0.03$ \\
                & 10450 & $ 1.3 \pm 0.3$   &1.0    &  Effelsberg 100-m, this work& $ 0.35 \pm 0.09 $ & $ 1.0 \pm 0.3$ \\
\hline
\noalign{\smallskip}
   WNB 1851+5707&  38   & $ 1800\pm 818$ &0.818  &  8C, Hales et al. 1995& $ 114\pm 11.9$  & $ 1686\pm 818 $ \\
                &  74   & $ 1240\pm 150$ &1.0    &  VLSS, Cohen et al. 2007& $ 65.7\pm 5.7 $ & $ 1174\pm 150$ \\
                &  151  & $ 723\pm 72$   &0.904  &  6CV, Hales et al. 1993& $ 36.3\pm 2.4$  & $ 687\pm 72$ \\
                &  232  & $ 530\pm 50$   &1.0    &  MIYUN, Zhang et al. 1997& $ 25.5\pm 1.4$  & $ 505\pm 50$ \\
                &  325  & $ 387\pm 4.4$  &1.0    &  WENSS, Rengelink et al. 1997& $ 19.3\pm 0.9$  & $ 368\pm 5$ \\
                &  365  & $ 437\pm 41$   &1.041  &  TEXAS, Douglas et al. 1996& $ 51.8\pm 1.8$  & $ 385\pm 41$ \\
                &  408  & $ 270\pm 10$   &1.091  &  B3.3, Altieri et al. 1999& $ 16.0\pm 0.7$  & $ 254\pm 10$ \\
                &  1400 & $ 53\pm 2.3$   &1.0    &  NVSS, Condon et al. 1998& $ 5.78\pm 0.12$ & $ 47\pm 2.3$ \\
                &  1425 & $ 41.9\pm 0.1$ &1.0    &  VLA, this work& $ - $           & $ 41.9\pm 0.1$ \\
                &  4850 & $ 6.1\pm 1.0$  &1.0    &  Effelsberg 100-m, this work& $ 2.09\pm 0.09 $& $ 4.0\pm 1.0$ \\
                &  4850 & $ 2.85\pm 0.03$&1.0    &  VLA, this work& $ - $           & $ 2.85\pm 0.03$ \\
                & 10450 & $ 1.2\pm 0.3$  &1.0    &  Effelsberg 100-m, this work& $ 1.12\pm 0.08 $& $ <1.0 $ \\
\hline
\noalign{\smallskip}
   B2  0120+33  &  74   & $2847\pm200$ &1.0   &  VLSS image, this work& $ -$                 & $ 2847\pm 200$ \\
                &  151  & $1690\pm170$ &0.904 &  6CVI, Hales et al 1993 &  $ 96.3\pm3.6$     & $ 1594\pm 170$ \\
                &  235  & $1100\pm88$  &1.0   &  GMRT, Giacintucci et al. in prep.&  $ -$        & $ 1100\pm88$ \\
                &  324  & $727\pm11.7$  &1.0   &  VLA, this work&  $ -$                       & $ 727\pm 11.7$ \\
                &  325  & $715\pm16$  &1.0   &  WENSS image, this work& $ -$                & $ 715\pm 16$ \\
                &  408  & $660\pm 20$  &1.0   &  Feretti \& Giovannini 1980&  $ 50.5\pm2.5$  & $ 609\pm 20$ \\
                &  610  & $372\pm18.6$  &1.0   &  GMRT, Giacintucci et al. in prep.&  $ -$        & $ 372\pm18.6$ \\
                &  1400 & $108\pm0.5$   &1.0   &  NVSS image, this work& $ -$                 & $ 108\pm 0.5$ \\
                &  1400 & $99.5\pm1.5 $   &1.0   &  VLA, this work&  $ -$                       & $ 99.5\pm 1.5$ \\
                &  2380 & $35\pm3.5 $  &1.0   &  ARECIBO, Dressel \& Condon 1978&  $ -$      & $ 35.0\pm 3.5$ \\
                &  4860 & $8.6\pm1.1$ &1.0   &  VLA, this work&  $ -$                       & $ 8.6\pm 1.1$ \\
                &  8460 & $<3.1 $       &1.0   &  VLA, this work&  $ -$                      & $ <3.1 $ \\
\hline
\noalign{\smallskip}
   B2  1610+29  &  74   & $1055\pm157$ &1.0   &  VLSS image, this work& $ -$  & $ 1055\pm 157$ \\
                &  151  & $774\pm9.5$  &1.237 &  7C, Riley et al. 1989& $ -$  & $ 774\pm 9.5$ \\
                &  325  & $412\pm5.2$  &1.0   &  WENSS, Rengelink et al. 1997& $ -$  & $ 412\pm 5.2$ \\
                &  408  & $349\pm 35$  &1.091 &  B2, Colla et al 1970& $ -$  & $ 349\pm 35$ \\
                &  610  & $240\pm 12$  &1.0   &  GMRT, Giacintucci et al. 2007& $ -$  & $ 240\pm 12$ \\
                &  1400 & $110\pm11$   &1.0   &  NVSS image, this work& $ -$  & $ 110\pm 11$ \\
                &  1452 & $101\pm2 $   &1.0   &  VLA, this work& $ -$  & $ 101\pm 2$ \\
                &  1502 & $98.8\pm2 $  &1.0   &  VLA, this work& $ -$  & $ 98.8\pm 2$ \\
                &  2639 & $49.8\pm3 $  &1.0   &  Effelsberg 100-m, this work& $ -$  & $ 49.8\pm 3$ \\
                &  4710 & $21.4\pm0.3$ &1.0   &  VLA, this work& $ -$  & $ 21.4\pm 0.3$ \\
                &  4850 & $19.0\pm0.4$ &1.0   &  Effelsberg 100-m, this work& $ -$  & $ 19.0\pm 0.4$ \\
                &  8350 & $4.8\pm0.6 $ &1.0   &  Effelsberg 100-m, this work& $ -$  & $4.8\pm0.6  $ \\
\hline
\end{tabular}
\end{center}
\label{integratedspectra}
\end{table*}

\section{MERLIN observations and data reduction}
Among the five dying sources, WNB1851+5707a is the most compact and bright.  We observed this source
 with the MERLIN interferometer in order to unveil the possible presence of a core or jets 
 unresolved by the VLA images.
WNB1851+5707a was observed in three days on April 2005 for a total on-source time
 of about 18 h with MERLIN (6 antennas) and Lovell. The observing frequency was
1.408 GHz, with a bandwidth of 16 MHz in both circular 
polarizations; the data were taken in spectral-line mode (32 $\times$ 0.5--MHz channels). 
OQ208 (1.00 Jy) and 3C286 (14.79 Jy) were used as flux calibration sources, 
while 1851+609 (0.25 Jy) as a phase calibrator.
The 1.4 GHz data from MERLIN and the VLA A-array were combined to increase the
 brightness sensitivity and uv-coverage, in order to image finer details of the
components seen in the L-Band MERLIN images, whilst also imaging the more diffuse 
radio emission that is observed in the VLA images. 
The length of the shortest baseline of the MERLIN array at 1.4 GHz is 6 km. At this frequency 
the interferometer is then not sensitive to structures larger than $\mathrm{\vartheta_{max}^{1.6 GHz} = 4\arcsec}$ 
(8.5 kpc at the distance of WNB1851+5707a). The combination of the 1.4 GHz MERLIN data with the VLA A-array 
data gives a shortest baseline of 0.68 km, so that $\mathrm{\vartheta_{max}^{1.4GHz}}$ = $\mathrm{64\arcsec}$ (136 kpc).
Several images were produced using the AIPS task IMAGR, and deconvolved with the
 multi-scale CLEAN algorithm. 
The full-resolution $0\farcs16$ MERLIN image is shown in the right panel of Fig.\,6. The only detected feature is a faint point-like
source with a flux density of $0.4\pm0.1$ mJy at the center of the optical galaxy. There is no evidence of active radio jets. Indeed,
 most of the flux density of  WNB1851+5707a comes from the extended structure already mapped in the VLA observations,
 confirming that WNB1851+5707a is a ``de-energized'' radio source. The MERLIN image is presented in combination with 
the VLA A-array at a resolution of $0\farcs4$ in the left panel of Fig.\,6. The MERLIN+VLA contours reveal that WNB1851+5707a is an amorphous radio source 
with a linear size of roughly $30$ kpc $\times$ $15$ kpc. The radio source is also considerably asymmetric being much more extended to the south of the optical galaxy.

\section{Analysis of the radio source spectra}
\begin{table*}[t]
\caption[]{Physical parameters of the dying sources. Uncertainties on fit parameters are at 1-$\sigma$ level.}
\label{fluxes}
\begin{center}
\begin{tabular}{cccccccccc}
\hline
\noalign{\smallskip}
   Source       & $\alpha_{\rm inj}$  & $\nu_{\rm b}$         & $t_{\rm OFF}/t_{\rm s}$ & $L_{151}$   & $B_{\rm min}$ & $u_{\rm min}$     &$t_{\rm s}$ & $t_{\rm CI}$ &  $t_{\rm OFF}$  \\
                &                     & MHz                   &                         & $\rm W/Hz$& $\rm \mu G$ & $\rm erg/cm^3$ & Myrs        & Myrs         &  Myrs       \\
\noalign{\smallskip}
\hline
\noalign{\medskip}
   WNB1734+6407 & $0.9_{-0.2}^{+0.3}$ & $230_{-200}^{+1550}$  & $0.3_{-0.2}^{+0.5}$ &   $2.6\times10^{25}$  & 10     &    $8.3\times10^{-12}$   & 86      &   60    &  26     \\

\noalign{\medskip}
\hline
\noalign{\medskip}

   WNB1829+6911 & $0.7_{-0.2}^{+0.6}$ & $50_{-20}^{+1500}$    & --                  &   $7.6\times10^{25}$   & 7.8     &   $5.3\times10^{-12}$   & 218        &   --     &  --   \\              
\noalign{\medskip}
\hline
\noalign{\medskip}
   WNB1851+5707a & $0.5_{-0.2}^{+0.2}$ & $220_{-170}^{+620}$   & $0.2_{-0.1}^{+0.2}$ &  $1.8\times10^{25}$   &  9.8   &    $8.9\times10^{-12}$   & 90        &  72     &  18   \\

\noalign{\medskip}
\hline
\noalign{\medskip}
   B2 0120+33 Global spec.& $0.6$ (fixed) & $440_{-140}^{+210}$    &$0.3_{-0.1}^{+0.1}$       &   $1\times 10^{24}$   & 5.2   &   $2.4\times10^{-12}$   & 141        &   99    &  42   \\
    Sp. index profile& $0.65$ (fixed) & $1840_{-380}^{+480}$      &$>0.8$                    &          &      &          & 69        &   $<13$      &  $>55$  \\

\noalign{\medskip}
\hline
\noalign{\medskip}
   B2 1610+29 Global spec.& $0.64_{-0.06}^{+0.06}$ & $5200_{-3400}^{+1200}$    &$>0.39$       &   $1.4\times10^{24}$   & 3.2   &   $9.3\times10^{-13}$   & 55        &   $<$33     &  $>$22   \\
    Sp. index profile& $0.6$ (fixed) & $4770_{-500}^{+600}$    &$0.8_{-0.1}^{+0.1}$        &                        &       &                         & 58        &   12        &  46  \\        
\noalign{\medskip}
\hline
\end{tabular}
\end{center}
\label{phypar}
\end{table*}

\subsection{Compilation of integrated spectra}
In addition to our own observations, we collected all the spectral information available in the literature for the
five dying radio galaxies. We made use of the CATS 
(the on-line Astrophysical CATalogs support System, at http://cats.sao.ru/; Verkhodanov et al. 1997)
to recover data from catalogs at different frequencies. We put all the flux densities from all these
reference on the absolute flux density of Baars et al. (1977) by scaling for the multiplicative factor listed 
in Helmboldt et al. (2008). The resulting catalogued flux densities ($S_{\nu}^{\rm catalogued}$) are given in Col.\,3 of 
Tab.\,5. The various catalogs we used to compile the integrated spectra have quite different angular resolution.
It is therefore necessary to remove the flux density due to confusing sources in order to recover a reliable spectrum
 for each dying source. We estimated the contribute of the confusing sources at each frequency on the basis of the VLA
images at 1.4 and 4.8\,GHz. The confusion flux density at a given frequency is given by the sum
\begin{equation}
S_{\nu}^{\rm confusion}=\sum_{i=1}^{N_{\nu}} S_{1425}^{i}(\nu/1425)^{-\alpha_{i}}
\end{equation}
where $\nu$ is the frequency in MHz, $S_{1425}$ is the flux density measured in the VLA image at 1425 MHz, $\alpha_{i}$ is the 
spectral index of the confusing source measured from the VLA images at 1.4 and 4.8\,GHz, and $N_{\nu}$ is the number
 of sources that fall within the beam of the catalog at frequency $\nu$.
Finally, we calculated the source flux density at a given frequency as
\begin{equation}
S_{\nu}=S_{\nu}^{\rm catalogued}-S_{\nu}^{\rm confusion}
\end{equation}
this is the quantity listed in Col.\,7 of Tab.\,5.

The total spectra of the five dying radio sources are show in bottom panel of Figs.\,1 to 5.
For four of them, namely WNB1734+6407, WNB1851+5707, B2 0120+33, and B2 1610+29, the integrated spectrum 
presents a strong  exponential cutoff in the observed frequency range. 
However, WNB 1829+6911 shows an evident flattening of the integrated spectrum at high frequency. This flattening is due to the
presence of the core-jet component that dominates the spectrum at the highest frequencies.

\subsection{Spectral modeling of the integrated spectra}

We modeled the integrated spectra assuming the radiative energy losses to be dominant with respect to other 
processes (e.g. adiabatic losses).
The pitch angles of the radiating electrons are assumed to be continually isotropized in a time that is shorter 
than the radiative time scale.
According to this assumption the synchrotron energy losses are statistically the same for all electrons.
After its birth the source is supposed to be fuelled at a constant rate (i.e. {\it the continuous injection phase}) 
by the nuclear activity, for a duration $t_{\rm CI}$. 
The injected  particles are assumed to have a power law energy spectrum $N(\epsilon) \propto \epsilon^{-\delta_{\rm inj}}$, which 
will result in a power law radiation spectrum 
with spectral index  $\alpha_{\rm inj} = (\delta_{\rm inj} - 1)/2$.
In this phase the source radio spectrum changes as a function of time in a way described by the shift of break frequency 
$\nu_{\rm b}$ to ever lower values as the time, $t_s$, increases:

\begin{equation}
\nu_{\rm b}\propto \frac{B}{(B^2+B_{\rm IC}^2)^2 t_{\rm s}^2}
\label{vb}
\end{equation} 
where $B$ and $B_{IC}=3.25(1+z)^2$ are the source magnetic field and the inverse Compton
equivalent magnetic field, respectively. Below and above $\nu_{\rm b}$ the spectral indices are
respectively $\alpha_{\rm inj}$ and $\alpha_{\rm inj}$+0.5.

At the time $t_{\rm CI}$ the power supply from the nucleus is switched-off. 
After that a new phase of duration $t_{\rm OFF}$ begins (i.e. the {\it  dying phase}). A new break 
frequency $\nu_{b\,high}$ then appears, beyond which the radiation
spectrum drops exponentially. This second high frequency break is
 related to the first by:

\begin{equation}
\nu_{b\,high}=\nu_{\rm b} \left(\frac{t_{\rm s}}{t_{\rm OFF}}\right)^{2}
\label{vb1vb2}
\end{equation}
 
where $t_{\rm s}=t_{\rm CI}+t_{\rm OFF}$ is the total source age (see e.g. Komissarov \&
 Gubanov 1994, Slee et al. 2001, Parma et al. 2007).

Thus, the above synchrotron model (hereafter CI$_{\rm OFF}$) is described by four parameters:
\begin{itemize}
\item[i)] $\alpha_{\rm inj}$, the injection spectral index;
\item[ii)] $\nu_{\rm b}$, the lowest break frequency;
\item[iii)] $t_{\rm OFF}/t_{\rm s}$, the dying to total source age ratio;
\item[iv)] $norm$, the flux normalization.
\end{itemize}

In the CI$_{\rm OFF}$ model the magnetic field strength is assumed to be uniform within the source. 

The fit of the  CI$_{\rm OFF}$ is shown as as line in bottom panels of Figs.\,1 to 5 while the best fit parameters
 are listed in Tab\,6. The fits are very good for all the five dying sources.

\begin{itemize}
\item[$-$]{In the case of  WNB1734+6407 the overall radio spectrum is particularly steep with an exponential cut-off beyond a frequency
 of about 230\,MHz. The spectral fit indicates that the source spent in the dying phase about 30\% of its total age ($t_{\rm OFF}/t_{\rm s}=0.3$).}

\item[$-$]{The source WNB 1829+6911 is a different case. Its low frequency spectrum is rather steep, but strongly flattens above 1.4 GHz, due to the presence of a 
bright core that is clearly detected in the 4.8\,GHz images. The core have a 
 flat spectrum characterized by $\alpha = 0.6$ between these two frequencies and is likely still active.
In order to account for this component in the spectral fit, we added a power law to the CI$_{\rm OFF}$ model, using the observed 
spectral index and flux density as the normalization. In this case we may be observing fading lobes 
(produced by a previous duty cycle), in conjunction with restarting activity in the core. However, we cannot constraint from the 
spectral fit the relative duration of the dying phase since the rising of the new flat spectrum core canceled out the
 the second high break frequency $\nu_{b\,high}$ of the fossil lobes.}

\item[$-$]{Both dying sources $a$ and $b$ contribute to the total spectrum of WNB 1851+5707. However, WNB 1851+5707a is ten times
brighter of WNB 1851+5707b (see Tab. 4). Indeed, if the radio sources have a similarly curved spectral shape, we can reasonably assume that the total spectrum
of WNB 1851+5707 is representative of source $a$ alone. The radio spectrum of WNB 1851+5707a is consistent with an injection 
spectral index of about $\alpha_{\rm inj}=0.5$ and cut-off almost exponentially beyond a break frequency of 220\,MHz. The duration of the 
dying phase is 20\% of the total source age. For what concerns WNB 1851+5707b we do not have the possibility to measure its complete total radio
 spectrum, given its closeness to its brighter companion. We can just report the particularly steep spectral index of the two fading lobes as seen in our VLA 
 images: $\alpha_{1.4}^{4.8}\simeq 2.3$.}

\item[$-$]{We fitted the integrated spectrum of B2 0120+33 with the CI$_{\rm OFF}$ model plus a power law which accounts for the emission of the radio core. The radio core however is
 very faint and its emission becomes important only at frequencies greater than 10\,GHz. At lower frequencies the source spectrum is dominated by the fading radio lobes and it 
can it be modeled with an injection spectral index of $\alpha_{\rm inj}=0.6$ and a break frequency of about 440\,MHz. According to the spectral fit the duration of
the dying phase should be about 30\% the source total age.}

\item[$-$]{Finally, the case of B2 1610+29. The source has an injection spectral index of  $\alpha_{\rm inj}=0.6$ and a break frequency somewhat higher with respect to the 
 other dying source presented above, $\nu_{b}\simeq 5200$\,MHz. Due to the lack of spectral information at frequencies higher that 10\,GHz, we can only put a
 lower limit on the duration of the dying phase: $t_{\rm OFF}/t_{\rm s}>0.4$. However, for this source we have a very good spatially resolved spectral index 
image and we can try to get some additional constraints on the source age from the fit of the spectral index profile.}

\end{itemize}

\subsection{Modeling of the spectral index profile in B2 0120+33 and B2 1610+29}
The two dying galaxies B2 0120+33 and B2 1610+29 are extended enough to permit a study of the spectral variation along the fading lobes.

For B2 0120+33 the profile of the spectral index between 325\,MHz and 1.4 GHz can be traced all along the lobes, see Fig.\,7. 
The spectral index trend has been obtained by averaging the flux densities at the 
two frequencies in circular regions of $36\arcsec$ in diameter centered as shown in the inset of Fig.\,7. The regions are much larger than the beam so that the spectral index measurements are effectively independent. The observed spectral index trend is rather smooth with  $\alpha_{0.3}^{1.4}$ close to about 1.2 over the whole radio source. Assuming a constant source expansion velocity the spectral index at a given distance from the core can
 be related to the synchrotron age of the electrons at that location. We assume that in B2 0120+33 the youngest electrons were injected by the jets close to the core during the active phase. This behavior is the typical one we observe in many FRI-type tailed radio sources (see e.g. Parma et al. 1999). In this case, we expect that the break frequency
scales as:

\begin{figure*}
\includegraphics[width=18cm]{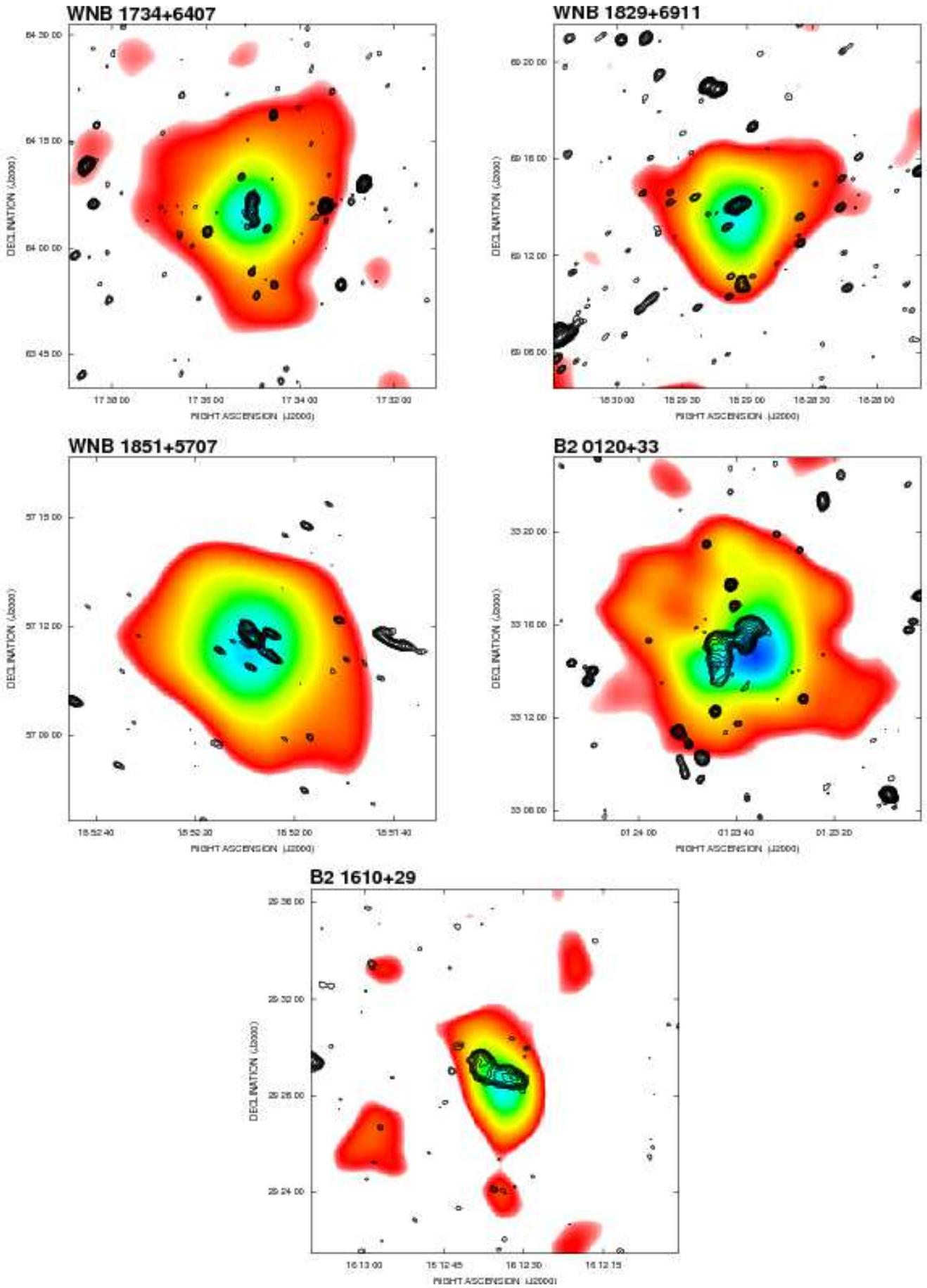}
\caption{RASS X-ray images with overlaid the VLA C-array contours at 1.4\,GHz. From top-left to bottom-right: Abell 2276, ZwCl 1829.3+6912, RXC J1852.1+5711, ZwCl 0107.5+3212, and Abell 2162. Radio contours start from a 3$\sigma$ rms level and increase by a factor of $\sqrt{2}$.}
\label{RASS}
\end{figure*}

\begin{equation}
\nu_{b}(d)=\frac{\nu_{b}(d_{\rm max})}{[(d/d_{\rm max})\cdot(1-t_{\rm OFF}/t_{s})+t_{\rm OFF}/t_{s}] ^2}
\label{vbdist}
\end{equation}

where the distance $d$ ranges from $d=0$ at the location of the galaxy, up to  $d=d_{\rm max}$ at
  the edge of each lobe. Thus, the break frequency along the lobes varies from a minimum of $\nu_{b}$, at $d=d_{\rm max}$,
 up to a maximum of $\nu_{\rm b\,high}$, at $d=0$. The two limiting break frequencies $\nu_{\rm b}$ and  $\nu_{\rm b\, high}$
 are exactly the same as given in Eqs.\, (\ref{vb}) and  (\ref{vb1vb2}), respectively. In Fig.\,7 we report the fit to the observed trend
 obtained in the case of $\alpha_{\rm inj}=0.65$ for different values of $t_{\rm OFF}/t_{\rm s}$. It is clear that low values of 
 $t_{\rm OFF}/t_{\rm s}$ correspond to spectral index gradients which are too steep if compared to the observed one. 
As $t_{\rm OFF}/t_{\rm s}$ increases, however, the expected spectral index trends get flatter and flatter and a best fit is 
found for  $t_{\rm OFF}/t_{\rm s}\ge0.8$. 

In the case of B2 1610+29 the profile of the 1.4 to 4.8 GHz spectral index can be traced all along the lobes, see Fig.\,8. The spectral index trend 
has been obtained by averaging the flux densities at the 
two frequencies in boxes perpendicular to the source major axis. The boxes are wide as the beam so that the spectral 
index measurements are effectively independent. The observed spectral index trend is rather smooth with  $\alpha_{1.4}^{4.8}$ ranging from about 1.2 at the edge of the lobes up to about 1.4 at the center of the radio source.
In  this particular case, we assume that the electrons were injected by the jets close to the end of each of the two lobes during the active phase. This
 spectral behaviour is common in FRII type radio sources but is also found in some FRI type radio sources (see Parma et al. 1999). In this case, we expect that the break frequency scales as:

\begin{equation}
\nu_{b}(d)=\frac{\nu_{b}(d_{\rm max})}{\{[(d_{\rm max}-d)/d_{\rm max}]\cdot(1-t_{\rm OFF}/t_{s})+t_{\rm OFF}/t_{s}\} ^2}
\label{vbdist}
\end{equation}

where the distance $d$ ranges from $d=0$ at the location of the galaxy, up to  $d=d_{\rm max}$ at
  the edge of each lobe. Thus, the break frequency along the lobes varies from a minimum of $\nu_{b}$, at $d=0$,
 down to a maximum of $\nu_{\rm b\,high}$, at $d=d_{\rm max}$. In Fig.\,8 we report the fit to the observed trend
 obtained in the case of $\alpha_{\rm inj}=0.6$ for different values of $t_{\rm OFF}/t_{\rm s}$. It is clear that low values of 
 $t_{\rm OFF}/t_{\rm s}$ correspond to spectral index gradients which are too steep if compared to the observed one. 
As $t_{\rm OFF}/t_{\rm s}$ increases, however, the expected spectral index trends get flatter and flatter and a best fit is 
found for  $t_{\rm OFF}/t_{\rm s}=0.8\pm0.1$. 

The behavior of the spectral index seen in B2 0120+33 and B2 1610+29 could be typical of many dying sources.
In practice, as a result of the jets switch-off, any pre-existent spectral index gradient along the lobes is quickly 
canceled since the break frequency reaches roughly 
the same value in each part of the source. We can therefore expect that extreme dying sources, those for which $t_{\rm OFF}/t_{\rm s}\simeq 1$, are characterized by very uniform spectral index distributions along the fading lobes. As the dying source gets older, the spectral index increases systematically but with small variations from point to point.

\subsection{Equipartition parameters}
We performed the minimum energy calculation for the five dying galaxies by considering the source power at 151 MHz, where the 
energy losses of the synchrotron electrons are less dramatic, and the volumes as measured from the VLA images at arcsecond resolution.
In the calculation of the sources power we made use of the 151 MHz flux density after the deconvolution of confusing sources. 
The resulting monochromatic powers in the sources rest frame, $L_{151}$, are listed in Tab.\,6 and place them below the FRI-FRII 
division (Fanaroff \& Riley 1974). 

We assume that the radio sources contain relativistic particles and magnetic fields uniformly distributed and in energy equipartition 
conditions. The equipartition parameters (magnetic field $B_{\rm min}$ and  energy density $u_{\rm min}$) are generally 
computed assuming that the relativistic particle energies are confined between a minimum $\epsilon_{\rm low}$ and a maximum $\epsilon_{\rm high}$,
corresponding to the observable radio frequency range, typically 10 MHz - 100 GHz (see, e.g., Pacholczyk 1970). 
This choice minimizes the source energetics required by the observed radiation in the radio band. However, a fixed frequency range 
corresponds to an energy range that depends on the source magnetic field, which may change from source to source. 
A fixed frequency range computation would miss the contribution from lower energy electrons, since the $\epsilon_{low}$ corresponding to 10 MHz is 
larger than 200 MeV for $B_{\rm min} \le 30~\mu G$ (Brunetti et al. 1997 and Beck \& Krause 2005). Because of this we have computed the equipartition parameters 
assuming a fixed low energy cutoff $\epsilon_{\rm low}$ = 10 MeV. The high energy  $\epsilon_{\rm high}$ cut-off is chosen to match the spectral break in the
 emission spectrum, $\nu_{\rm b}$, since no electrons are present in the source beyond this limit, see Parma et al. (2007) for further details. 

The equipartition parameters computed in this way are reported in Tab. 6. A comparison with fixed frequency range 
calculations shows that the values of our $B_{\rm min}$ are larger by up to a factor two.

\subsection{Radiative ages}

Assuming a constant magnetic field and neglecting expansion losses, the total source age can be calculated from the break
frequency, $\nu_{b}$:

\begin{equation}
t_{\rm s}= 1590 \frac{B^{0.5}}{(B^2+B_{\rm IC}^2) [(1+z)\nu_{\rm b}]^{0.5}}
\label{synage}
\end{equation}

where the synchrotron age $t_{\rm s}$ is in Myr, the magnetic field in $\mu$G, the break 
frequency $\nu_{\rm b}$ in GHz, while the inverse Compton equivalent field (see Sect. 5.2).

By adopting the equipartition value $B_{\rm min}$ for the magnetic field strength we can thus
derive the synchrotron age $t_{\rm s}$. Finally, from the ratio $t_{\rm OFF}/t_{\rm s}$, which
is also given by the fit, we can determine the absolute durations of the active
and dying phases, $t_{\rm CI}$  and $t_{\rm OFF}$. The {\it total} ages of the five dying sources are of the order 
of $10^{8}$ yrs (with uncertainties up to 50\%). For all sources but  WNB 1829+6911 we are also able to estimate the absolute duration of the active and dying 
phases that result in the range $10^7~ -~ 10^8$ yrs. We note that these numbers should be considered us upper limits to the real source
 ages. It is likely that expansion losses may have played a role during the early stages of the radio source growth, and,
 in the same manner, the magnetic field strength was presumably higher than the actual values. Considering both these effects,
 the shift of $\nu_{\rm b}$ to low frequencies may have proceeded much rapidly than expected by the simple radiative model we considered, and hence,
 the ages would have been overestimated. 

\section{The X-ray environment}
The gaseous environment in which radio galaxies are embedded may play a fundamental role in the latest stages of
the radio source life. Although no firm conclusions can be drawn about the environment of dying sources, due to
the small number of objects involved, it seems that there is a tendency for dying sources to reside in dense environments.
Parma et al. (2007) found that about half of the dying sources of their sample are located in clusters of galaxies while only a few
appear to be isolated.

We have made a search in the RASS for possible X-ray counterparts of our five dying radio galaxies. 
We extracted the 0.1 - 2.4 keV image of a $15\arcmin \times 15\arcmin$ field around
each radio source. The X-ray images were corrected for the background and smoothed with a $\sigma= 45\arcsec$ Gaussian kernel.
The RASS count-rate images are shown in Figs.\,9 along with the 1.4\,GHz VLA C-array iso-contours overlaid. The five dying galaxies presented here do not represent 
an exception to the rule: each source is located, at least in projection, at 
the center of an X-ray emitting cluster or galaxy group. The clusters of galaxies are Abell 2276, ZwCl 1829.3+6912, 
RX J1852.1+5711, ZwCl 0107.5+3212, and Abell 2162 for the radio 
sources WNB 1734+6407, WNB 1829+6911, WNB 1851+5707, B2 0120+33, and B2 1610+29, respectively. 

\subsection{Abell 2276}
Very little is known about this cluster of galaxies besides that 
Jones \& Forman (1999) report an X-ray luminosity $1.02 \times 10^{44}$ erg/s in the 0.5-4.5 keV band based on {\it Einstein} data (all X-ray
 luminosities have been rescaled to our adopted cosmology).
The dying source WNB 1734+6407 is associated with the central galaxy at the peak of the X-ray emission, see top-left panel in Fig.\,9. 

\subsection{ZwCl 1829.3+6912}
This galaxy cluster is also known as MACS J1829.0+6913 or RX J1829.0+6912. Only sparse information are 
present in literature for this cluster too. It appears as catalog number NEP5700 in the ROSAT North Ecliptic Pole survey by Gioia et al. (2003). These
authors report an X-ray luminosity of $1.06 \times 10^{44}$ erg/s in the 0.5-2.0 keV band. WNB 1829+6911 associated to the cD galaxy located close to the 
X-ray peak, see top-right panel in Fig.\,9. 

\subsection{RX J1852.1+5711}
The extended ROSAT Bright Cluster Sample (Ebeling et al. 2000) lists an intrinsic X-ray luminosity in the 0.1-2.4 keV band of $0.95\times 10^{44}$ erg/s
for this cluster. The dying sources WNB 1851+5707a is associated to the galaxy at the center of the X-ray emission while the dying source WNB 1851+5707b 
and its host galaxy lie 30\arcsec to the north east. WNB 1851+5707a and WNB 1851+5707a appear blended together at the resolution of the VLA C-array
shown as iso-contours in the middle-left panel of Fig.\,9.

\subsection{ZwCl 0107.5+3212}
The dying source B2 0120+33 is identified with the galaxy NGC 507 at the center of the Zwicky cluster
0107.5+3212 and is one of the brightest galaxies in a very dense region. There are 80 galaxies with magnitude $m_{pg} < 15.7$ within a projected
distance of 1.5 Mpc from NGC 507. This is the so-called Pisces Group (Parma et al. 1986). The X-ray emission has been studied in detail
by many authors with different satellites, see e.g. Kraft et al. (2004) and references therein.
NGC 507 is surrounded by an irregular large-scale X-ray emission with a peak coincident with the position of the optical galaxy, and several secondary
peaks, due to the interaction between the radio lobes and the surrounding gas (Paolillo 2003). Kim \& Fabbiano (1995) report an X-ray luminosity of 
$0.8 \times 10^{43}$ in the 0.1-2.4 keV band for the NGC 507 group. Kraft et al. (2004) observed a sharp edge in the X-ray surface
brightness profile in a Chandra observation of NGC 507. These authors suggested that the discontinuity is likely caused by an elemental 
abundance jump driven by the inflation of the radio lobes and entrainment of material from the central regions of the cluster.

\subsection{Abell 2162}
Abell 2162 is member of the Hercules super cluster complex (Einasto et al. 2001). The X-ray emission of this nearby cluster of galaxies has been 
studied by Ledlow et al. (2003) who report an X-ray luminosity in the 0.5-2.0 keV band of $0.16\times 10^{43}$ erg/s based on ROSAT data. 
The fading radio lobes of B2 1610+29 are associated to NGC 6086, the brightest galaxy of the cluster located close to the X-ray peak (Fig.\,9 bottom panel).\\

Indeed, these results suggest that the occurrence of dying sources could be higher in galaxy clusters, confirming the findings of Parma et al. (2007).

In order to determine if the environment really plays a role in increasing the occurence or the duration of the dying phase we need
to estimate the relative fraction of dying sources in isolated and cluster galaxies. This is not trivial because estimating the completeness
of the inhomogeneous  radio, optical, and X-ray samples is hard. 
However, by imposing safer selection limits we can attempt to perform some very rough statistical analysis.
A complete sample of 119 sources can be constructed from the minisurvey by considering all galaxies with optical magnitudes $m_r<16$ 
and radio flux density $S_{325\rm MHz}>30$ mJy (de Ruiter et al. 1998). If we further restrict our attention to the elliptical galaxies we have 
a complete sample of 90 radio sources. Note that we consider WNB 1851+5707 as two distinct radio galaxies, both in the dying phase.

For each of these radio sources we searched in the NED database for the presence of a galaxy cluster or
group to within $10\arcmin$ from the position of host galaxy. We found that 21 out of 90 sources lies, in projection, in Abell, Zwicky or
 X-ray selected galaxy clusters or groups, while we considered the rest to be isolated.
In this complete sample there are three of our dying sources: WNB 1734+6407, WNB 1851+5707a and WNB 1851+5707b. The restarting source WNB1829+6911 exceeds the
optical limit due to its larger distance and therefore is not included in the complete sample.
Therefore, the fraction of dying source in general is $\sim 3.3$\% (3/90).  However, this fraction depends on the environment 
since the frequency of dying sources in clusters is 14\% (3/21).

In the complete B2 bright sample (Colla et al. 1975), we found the presence of fossil radio lobes in four of 54 elliptical galaxies. 
Of these, three (B2 0120+33, B2 1610+29, and B2  1636+39) are in cluster and only one (B2 0924+30) can be considered relatively isolated\footnote{
The dying source B2 0924+30 is member of the nearby poor cluster WBL224 (White et al. 1999), but no extended X-ray emission is present
in the RASS.}. With the same criterion describe above, we estimated that about 21 radio sources of the B2 sample are located in projection
close to the center of a galaxy cluster or group. The fraction of dying source in general is 7\% (4/54), while the fraction of dying sources in clusters
 is 14\% (3/21).

The fractions of dying sources (both in general and in cluster) in the minisurvey and in the B2 samples are consistent to within the large
uncertainties. We then merged the statistics of the two samples in order to estimate the probability for a dying source to be in cluster:

\begin{equation}
P(cluster\vert dying)= \frac{P(dying\vert cluster)P(cluster)}{P(dying)},
\end{equation}
where $P(dying\vert cluster)=6/42$ is the fraction of dying sources in cluster, $P(cluster)=42/144$ is the total fraction of radio sources in cluster,
 and $P(dying)=7/144$ is the total fraction of dying sources. Indeed, we obtained that the probability for a dying source to be in cluster
is $P(cluster\vert dying)=6/7$, i.e. $\sim 86$\%. We estimated through a simple Monte Carlo simulation that if the possibility to have a dying source were independent from the environment, the probability to have 6/7 of dying sources in cluster by chance is less than 0.5\%.

One possibility is that the low-frequency 
radio emission from the fading radio lobes last longer if their expansion is somewhat reduced or even stopped.   
Another possible explanation is that the frequency of dying sources in cluster is higher. Radio sources in dense environment, 
like e.g. the center of cooling core clusters, may have a peculiar accretion mode which results in a bursting duty cycle 
sequence of active and quiescent periods. Of the five dying galaxies presented in this work, WNB1829+6911 and B2 0120+33 show evidence 
for a flat-spectrum core. A very faint core is detected also in the MERLIN full-resolution image of WNB1851+5707a, 
while a steep spectrum feature is detected at the center of the host galaxy of WNB1734+6407. Thus, although the extended fading radio lobes in 
these sources are not powered by kpc-scale jets anymore, the AGN is still radio active at very
low levels.

To investigate these hypotheses we need to compare in detail the actual fading radio structures with the properties of the X-ray
emitting gas. We recently have had the chance to observe three more of these clusters, Abell 2276, 
ZwCl 1829.3+6912, and RXC J1852.1+5711, with the Chandra satellite. We report on the results of these observations on 
a forthcoming paper (Murgia et al., in preparation).

\section{Summary}
In this work we presented the study of five `dying' radio galaxies belonging to the Westerbork
Northern Sky Survey minisurvey and to the B2 bright sample: WNB1734+6407, WNB1829+6911, WNB1851+5707, B2 0120+33, and B2 1610+29.

These sources have been selected on the basis of their extremely steep broad-band radio spectra, which
is a strong indication that these objects either belong to the rare class of dying radio galaxies or 
that we are observing `fossil' radio plasma remaining from a previous nuclear activity. We derived the relative duration of 
the dying phase from the fit of a synchrotron radiative model to the radio spectra of the sources.
The modeling of the integrated spectra and the deep spectral index images obtained with the VLA 
confirmed that in these sources the 
central engine has ceased to be active for a significant fraction of their lifetime although their extended lobes have
not yet completely faded away. In the cases of  WNB1829+6911 and  B2 0120+33, the fossil radio lobes are seen in conjunction
with a currently active core. We found that WNB1851+5707 is in reality composed by two distinct dying sources that originally  appear blended together at the
 lower angular resolution of the WENSS. 

All the dying sources of our sample are located (at least in projection) at the
center of an X-ray emitting cluster.

Although no firm conclusions can be drawn due to the small number statistics involved,
our results suggest that the duration of the dying phase for a radio source in
clusters can be significantly higher with respect to that of a radio galaxy in the field.
The simplest interpretation for the tendency of dying galaxies to be found in cluster is that the low-frequency 
radio emission from the fading radio lobes last longer if their expansion is somewhat reduced or even stopped by
 the pressure of a particularly dense intra-cluster medium.  
Another possibility is that the occurrence of dying sources is higher in galaxy clusters.
Of the five dying galaxies, WNB1829+6911 and B2 0120+33 show evidence for a flat-spectrum core. A very faint
core is detected also in the MERLIN full-resolution image of WNB1851+5707a, while a steep spectrum feature 
is detected at the center of the host galaxy of WNB1734+6407. Thus, although the extended fading radio lobes in 
these sources are not powered by kpc-scale jets anymore, the AGN is still radio active at very
low levels. Indeed, this may suggest that radio sources in dense environment, like e.g. the center of cooling core clusters, 
have a peculiar accretion mode which results in a bursting duty cycle sequence of active and quiescent periods. 
This result could have important implications for theories
of the life cycles of radio sources and AGN feedback in clusters of galaxies but awaits
confirmation from future observations of larger, statistically significant, samples of objects.

We may expect the existence of a large population of dying radio sources that have been missed from the current surveys because of  
very steep spectra. These sources are very faint at centimeter wavelengths but should still be visible at frequency below 100 MHz if they are only subject
 to radiative losses. Due to their sensitivity and angular resolution the upcoming low-frequency radio interferometers (such as LOFAR and LWA and, in 
the next future, the SKA) represent the ideal instruments to discover and study in detail these elusive objects.

\begin{acknowledgements}
We are grateful to the anonymous referee for very useful comments that improved this paper.
FG and MM thank the hospitality of the Harvard-Smithsonian Center for Astrophysics where part of this
research was done.  Support was provided by Chandra grant GO9-0133X, NASA contract NAS8-39073, and the Smithsonian
Institution. We acknowledge financial contribution from the agreement ASI-INAF I/009/10/0.
The National Radio Astronomy Observatory is operated by Associated Universities, Inc., under contract with the National Science Foundation.
This research made use of the NASA/IPAC Extragalactic Database (NED) which is operated by the Jet Propulsion Laboratory, California Institute of Technology, under contract 
with the National Aeronautics and Space Administration.
MERLIN is a National Facility operated by the University of Manchester at Jodrell Bank Observatory on behalf of STFC.
This work has benefited from research funding from the European Community's sixth Framework Programme under RadioNet R113CT 2003 5058187.
This research made use also of the CATS Database (Astrophysical CATalogs support System).
The optical DSS2 red images were taken from: http://archive.eso.org/dss/dss.
We thank the staff of the National Galileo Telescope (TNG) for carrying out the
optical observations in service mode.
\end{acknowledgements}


\begin{thebibliography}{}
\bibitem[]{} Altieri, F., et al. 1999, Laurea Thesis, University of Bologna 
\bibitem[Baars et al.(1977)]{1977A&A....61...99B} Baars, J.~W.~M., Genzel, R., Pauliny-Toth, I.~I.~K., \& Witzel, A.\ 1977, \aap, 61, 99 
\bibitem[Beck \& Krause(2005)]{2005AN....326..414B} Beck, R., \& Krause, M.\ 2005, Astronomische Nachrichten, 326, 414 
\bibitem{} Blumenthal, G.~R., \& Gould, R.~J.\ 1970, Reviews of Modern Physics, 42, 237 
\bibitem[Brunetti et al.(1997)]{1997A&A...325..898B} Brunetti, G., Setti, G., \& Comastri, A.\ 1997, \aap, 325, 898 
\bibitem[Colla et al.(1970)]{1970A&AS....1..281C} Colla, G., Ficarra, A.; Formiggini, L., et al.\ 1970, \aaps, 1, 281 
\bibitem[Colla et al.(1972)]{1972A&AS....7....1C} Colla, G., et al.\ 1972, \aaps, 7, 1 
\bibitem[Colla et al.(1973)]{1973A&AS...11..291C} Colla, G., et al.\ 1973, \aaps, 11, 291 
\bibitem[Colla et al.(1975)]{1975A&A....38..209C} Colla, G., Fanti, C., Fanti, R., Gioia, I., Lari, C., Lequeux, J., Lucas, R., \& Ulrich, M.-H.\ 1975, \aap, 38, 209 
\bibitem[Cohen et al.(2007)]{2007AJ....134.1245C} Cohen, A.~S., Lane, W.~M., Cotton, W.~D., et al. \ 2007, \aj, 134, 1245 
\bibitem[1998]{Condon98}Condon, J.J., Cotton, W.D., Greisen, E.W. et al. 1998, AJ, 115, 1693
\bibitem[Cordey(1987)]{1987MNRAS.227..695C} Cordey, R.~A.\ 1987, \mnras, 227, 695 
\bibitem[Douglas et al.(1996)]{1996AJ....111.1945D} Douglas, J.~N., Bash, F.~N., Bozyan, F.~A., et al. \ 1996, \aj, 111, 1945 
\bibitem[Dressel \& Condon(1978)]{1978ApJS...36...53D} Dressel, L.~L., \& Condon, J.~J.\ 1978, \apjs, 36, 53 
\bibitem[Ebeling et al.(2000)]{2000MNRAS.318..333E} Ebeling, H., Edge, A.~C., Allen, S.~W., Crawford, C.~S., Fabian, A.~C., \& Huchra, J.~P.\ 2000, \mnras, 318, 333
\bibitem[Einasto et al.(2001)]{2001AJ....122.2222E} Einasto, M., Einasto, J., Tago, E., M{\"u}ller, V., \& Andernach, H.\ 2001, \aj, 122, 2222 
\bibitem[Fanaroff \& Riley(1974)]{1974MNRAS.167P..31F} Fanaroff, B.~L., \& Riley, J.~M.\ 1974, \mnras, 167, 31P 
\bibitem[Feretti \& Giovannini(1980)]{1980A&A....92..296F} Feretti, L., \& Giovannini, G.\ 1980, \aap, 92, 296 
\bibitem[Ferrari et al.(2008)]{2008SSRv..134...93F} Ferrari, C., Govoni, F., Schindler, et al. \ 2008, Space Science Reviews, 134, 93 
\bibitem[Gentile et al.(2007)]{2007ApJ...659..225G} Gentile, G., Rodr{\'{\i}}guez, C., Taylor, G.~B., et al.\ 2007, \apj, 659, 225 
\bibitem[Giacintucci et al.(2007)]{2007A&A...476...99G} Giacintucci, S., Venturi, T., Murgia, M., et al. \ 2007, \aap, 476, 99 
\bibitem[Gioia et al.(2003)]{2003ApJS..149...29G} Gioia, I.~M., Henry, J.~P., Mullis, et al. \ 2003, \apjs, 149, 29 
\bibitem[1988]{Giov88} Giovannini, G., Feretti, L., Gregorini, L., \& Parma, P.1988, A\&A, 199, 73
\bibitem[Helmboldt et al.(2008)]{2008ApJS..174..313H} Helmboldt, J.~F., Kassim, N.~E., Cohen, A.~S., et al. \ 2008, \apjs, 174, 313 
\bibitem[Hales et al.(1990)]{1990MNRAS.246..256H} Hales, S.~E.~G., Masson, C.~R., Warner, P.~J., \& Baldwin, J.~E.\ 1990, \mnras, 246, 256 
\bibitem[Hales et al.(1991)]{1991MNRAS.251...46H} Hales, S.~E.~G., Mayer, C.~J., Warner, P.~J., \& Baldwin, J.~E.\ 1991, \mnras, 251, 46 
\bibitem[Hales et al.(1993)]{1993MNRAS.263...25H} Hales, S.~E.~G., Baldwin, J.~E., \& Warner, P.~J.\ 1993, \mnras, 263, 25 
\bibitem[Hales et al.(1995)]{1995MNRAS.274..447H} Hales, S.~E.~G., Waldram, E.~M., Rees, N., \& Warner, P.~J.\ 1995, \mnras, 274, 447
\bibitem[2004]{Jamrozy04}Jamrozy, M., Klein, U., Mack, et al. \ 2004, A\&A, 427, 79
\bibitem[Jones \& Preston(2001)]{2001AJ....122.2940J} Jones, D.~L., \& Preston, R.~A.\ 2001, \aj, 122, 2940 
\bibitem[]{} Kim, D. W., \& Fabbiano, G. 1995, ApJ, 441, 182
\bibitem[Komissarov \& Gubanov(1994)]{1994A&A...285...27K} Komissarov, S.~S., \& Gubanov, A.~G.\ 1994, \aap, 285, 27
\bibitem[Kraft et al.(2004)]{2004ApJ...601..221K} Kraft, R.~P., Forman, W.~R., Churazov, E., Laslo, N., Jones, C., Markevitch, M., Murray, S.~S., \& Vikhlinin, A.\ 2004, \apj, 601, 221 
\bibitem[Ledlow et al.(2003)]{2003AJ....126.2740L} Ledlow, M.~J., Voges, W., Owen, F.~N., \& Burns, J.~O.\ 2003, \aj, 126, 2740 
\bibitem[Liuzzo et al.(2010)]{2010A&A...516A...1L} Liuzzo, E., Giovannini, G., Giroletti, M., \& Taylor, G.~B.\ 2010, \aap, 516, A1 
\bibitem[Murgia et al.(1999)]{1999A&A...345..769M} Murgia, M., Fanti, C., Fanti, R., et al. \ 1999, \aap, 345, 769 
\bibitem[Pacholczyk(1970)]{1970ranp.book.....P} Pacholczyk, A.~G.\ 1970,  Series of Books in Astronomy and Astrophysics, San Francisco: Freeman, 1970,  
\bibitem[Paolillo et al.(2003)]{2003ApJ...586..850P} Paolillo, M., Fabbiano, G., Peres, G., \& Kim, D.-W.\ 2003, \apj, 586, 850 
\bibitem[Parma et al.(1986)]{1986A&AS...64..135P} Parma, P., de Ruiter, H.~R., Fanti, C., \& Fanti, R.\ 1986, \aaps, 64, 135 
\bibitem[Parma et al.(1999)]{1999A&A...344....7P} Parma, P., Murgia, M., Morganti, R., Capetti, A., de Ruiter, H.~R., \& Fanti, R.\ 1999, \aap, 344, 7 
\bibitem[Parma et al.(2007)]{2007A&A...470..875P} Parma, P., Murgia, M., de Ruiter, et al. \ 2007, \aap, 470, 875 
\bibitem[1997]{Renge97} Rengelink, R.B., Tang, Y., de Bruyn, A.G. et al. 1997, A\&AS, 124, 259
\bibitem[Riley(1989)]{1989MNRAS.238.1055R} Riley, J.~M.\ 1989, \mnras, 238, 1055 
\bibitem[de Ruiter et al.(1998)]{1998A&A...339...34D} de Ruiter, H.~R., Parma, P., Stirpe, et al. \ 1998, \aap, 339, 34 
\bibitem{} Rybicki, G.~B., \& Lightman, A.~P.\ 1979, New York, Wiley-Interscience, 1979. 393 p.
\bibitem[Saripalli et al.(2002)]{2002ApJ...565..256S} Saripalli, L., Subrahmanyan, R., \& Udaya Shankar, N.\ 2002, \apj, 565, 256 
\bibitem[Saripalli et al.(2003)]{2003ApJ...590..181S} Saripalli, L., Subrahmanyan, R., \& Udaya Shankar, N.\ 2003, \apj, 590, 181 
\bibitem[Scheuer(1974)]{1974MNRAS.166..513S} Scheuer, P.~A.~G.\ 1974, \mnras, 166, 513 
\bibitem[Schoenmakers et al.(2000)]{2000MNRAS.315..395S} Schoenmakers, A.~P., de Bruyn, A.~G., R{\"o}ttgering, H.~J.~A., \& van der Laan, H.\ 2000, \mnras, 315, 395 
\bibitem[Slee et al.(2001)]{2001AJ....122.1172S} Slee, O.~B., Roy, A.~L., Murgia, M., et al. \ 2001, \aj, 122, 1172 
\bibitem[Sohn et al.(2003)]{2003A&A...404..133S} Sohn, B.~W., Klein, U., \& Mack, K.-H.\ 2003, \aap, 404, 133 
\bibitem[Verkhodanov et al.(1997)]{1997ASPC..125..322V} Verkhodanov, O.~V., Trushkin, S.~A., Andernach, H., \& Chernenkov, V.~N.\ 1997, Astronomical Data Analysis Software and Systems VI, 125, 322 
\bibitem[]{}White, R.~A., Bliton, M., Bhavsar, S.~P., Bornmann, P., Burns, J.~O., Ledlow, M.~J., \& Loken, C.\ 1999, \aj, 118, 2014 
\bibitem[Zhang et al.(1997)]{1997A&AS..121...59Z} Zhang, X., Zheng, Y., Chen, et al. \ 1997, \aaps, 121, 59 


\end{thebibliography}
\end{document}